\documentclass[12pt,a4paper]{article}
\usepackage{a4wide}
\usepackage{amsmath}
\usepackage{amssymb}
\usepackage{epsfig}
\usepackage{subfigure}
\usepackage{exscale}
\usepackage{float}
\usepackage{bbm}
\usepackage[numbers,sort&compress]{natbib}
\usepackage{pst-plot, pstricks,pst-math}
\usepackage{fancybox,amssymb,color}
\usepackage{graphicx}
\usepackage{pstricks, color, graphicx, epsfig, psfrag}
\usepackage{amsfonts,amsmath,amssymb,slashed}
\usepackage{dsfont}
\usepackage{bbm,bm}
\usepackage{fancyhdr, a4wide}
\usepackage[english]{babel}
\usepackage{subfigure}

\makeatletter
\@addtoreset{equation}{section}
\makeatother

\title{Spin Order and Entropy in Antiferromagnetic Films Subjected to Magnetic Fields}

\author{Christoph P.\ Hofmann$^a$ \\ \\
\normalsize{$^a$ Facultad de Ciencias, Universidad de Colima} \\
\vspace{0.3cm}
\normalsize{Bernal D\'iaz del Castillo 340, Colima C.P.\ 28045, Mexico} \\}

\begin{document}
\maketitle

\begin{abstract} \normalsize

Using systematic effective field theory, we explore the properties of antiferromagnetic films subjected to magnetic and staggered fields
that are either mutually aligned or mutually orthogonal. We provide low-temperature series for the entropy density in either case up to
two-loop order. Invoking staggered, uniform and sublattice magnetizations of the bipartite antiferromagnet, we investigate the subtle
order-disorder phenomena in the spin arrangement, induced by temperature, magnetic and staggered fields -- some of which are quite
counterintuitive. In the figures we focus on the spin-$\frac{1}{2}$ square-lattice antiferromagnet, but our results are valid for any other
bipartite two-dimensional lattice.

\end{abstract}

\maketitle

\section{Introduction}
\label{Intro}

The present work is part of an ongoing program the aim of which it is to systematically analyze the thermodynamic properties of
antiferromagnetic systems using magnon effective field theory. While three-dimensional antiferromagnets have been discussed within this
perspective in Refs.~\citep{HL90,Leu94a,Hof99a,Hof99b} -- and more recently in Refs.~\citep{Hof17b,BH17,BH19,Hof20c} -- here we continue
exploring antiferromagnetic films.

Earlier effective field theory based papers on antiferromagnetic films include Refs.~\citep{CHN89,Fis89,HL90,HN93,Hof10}. More recent
studies are provided by Refs.~\citep{Hof17,Hof20a,Hof20b}. In the present investigation we consider antiferromagnetic films subjected to
magnetic and staggered fields that are either mutually aligned or mutually orthogonal. A thorough and rigorous analysis of this situation --
in particular at the two-loop level that is our objective -- still appears to be lacking which motivates our work. New analytic results
comprise low-temperature series for the entropy density and the sublattice magnetizations of the bipartite two-dimensional antiferromagnet
up to two-loop order. As it turns out, these low-temperature representations are parameter free in the sense that they do not involve
next-to-leading order effective constants, but solely depend on the spin stiffness and the zero-temperature staggered magnetization, i.e.,
the order parameter.

It should be pointed out that the path we pursue here -- based on the systematic effective Lagrangian method -- is not the conventional
choice to address antiferromagnetic systems. Rather, more standard techniques to study antiferromagnetic films are modified spin-wave
theory \citep{Tak89,AS93,VCC05,SSKPWLB05,HSSK07,TZSS08,KSHK08,HRO10,SST11}, numerical simulations \citep{San99,SS02,PR15}, exact
diagonalization and other approaches \citep{Gho73,FKLM92,Glu93,MG94,SSS94,HSR04,CZ09,LL09,FZSR09,SST13}.

Our main theme is the exploration of order-disorder phenomena in the spin arrangement that are induced by temperature and by the magnetic
and staggered fields. To this end, referring specifically to the spin-$\frac{1}{2}$ square-lattice antiferromagnet, we provide plots for the
entropy density, as well as the uniform, staggered and sublattice magnetizations.

Before delving into the effects at finite temperature, we first survey the situation at $T$=0. In both cases -- mutually parallel and
orthogonal fields -- staggered, uniform and sublattice magnetizations grow when the magnetic or the staggered field become stronger, which
can be understood by suppression of quantum fluctuations.

We then analyze the behavior of the same observables at finite -- but first fixed -- temperature by varying magnetic and staggered field
strength. Regarding the entropy density we identify two opposite tendencies: (1) if the staggered field becomes stronger, the entropy
decreases, and (2) if the magnetic field becomes stronger, the entropy increases. Naively, the staggered field enhances spin order by
enforcing antialignment of the spins, whereas the magnetic field perturbs the antialigned array of spins. These observations that apply to
mutually parallel and mutually orthogonal fields alike, are confirmed by the finite-temperature behavior of the staggered, uniform and
sublattice magnetizations.

In both configurations of external fields a finite-temperature uniform magnetization is induced that initially even grows as temperature
rises -- this outcome is rather counterintuitive. It should be pointed out that such counterintuitive phenomena not only emerge in
bipartite antiferromagnetic films, but also have been observed in frustrated two-dimensional antiferromagents as well as in
antiferromagnetic spin chains and ladders.

While all effects have been illustrated by the spin-$\frac{1}{2}$ square-lattice antiferromagnet, any other bipartite two-dimensional
antiferromagnet behaves in a qualitatively similar way. Our results and the observed order-disorder phenomena are universal in this sense
-- microscopic details are merely encoded in the concrete values the spin stiffness and zero-temperature staggered magnetization.

The paper is organized as follows. In Sec.~\ref{MutuallyParallel}, after a few general remarks on antiferromagnetic films in magnetic
fields aligned with the order parameter, we derive the two-loop representation for the entropy density and examine its behavior in external
fields. In the same section we then investigate in detail how temperature, magnetic and staggered fields influence spin order by providing
various figures that include uniform, staggered and sublattice magnetizations for the spin-$\frac{1}{2}$ square-lattice antiferromagnet.
Some effects are quite counterintuitive. Along the same lines -- and by providing analogous figures -- in Sec.~\ref{MutuallyOrthogonal}, we
analyze antiferromagnetic films subjected to magnetic and staggered fields that are mutually orthogonal, and point out similarities and
differences with respect to the configuration of mutually parallel external fields. In Sec.~\ref{conclusions} we conclude. Finally, in
appendix \ref{appendixA} -- for self-consistency -- we list the relevant kinematical functions for antiferromagnets in mutually orthogonal
external fields.

\section{Antiferromagnetic Films in Mutually Parallel Staggered and Magnetic Fields}
\label{MutuallyParallel}

\subsection{Preliminaries}
\label{prelim1}

Within the microscopic point of view, antiferromagnetic films are captured by the quantum Heisenberg model that incorporates an external
magnetic (${\vec H}$) and a staggered (${\vec H_s}$) field as 
\begin{equation}
\label{HeisenbergZeemanH}
{\cal H} \, = \, - J \, \sum_{n.n.} {\vec S}_m \! \cdot {\vec S}_n \, - \, \sum_n {\vec S}_n \cdot {\vec H} \, - \, \sum_n (-1)^n {\vec S}_n
\! \cdot {\vec H_s} \, , \qquad J < 0 \, , \quad J = const. 
\end{equation}
The notation "n.n." indicates that the sum refers to nearest neighbor spins only. While our focus in the numerical analysis is
square-lattice geometry, we allow for a general bipartite lattice.

In the absence of external fields, two degenerate spin-wave branches (magnon modes) emerge, obeying the dispersion law
\begin{equation}
\label{disprelAF}
\omega(\vec k) \, = \, v|{\vec k}| + {\cal O}({\vec k}^3) \, , \qquad {\vec k} = (k_1,k_2) \, .
\end{equation}
The quantity $v$ is the spin-wave velocity. The specific configuration of external fields in the present section is
\begin{equation}
\label{externalFields}
{\vec H} = (H,0,0) \, , \qquad {\vec H}_s = (H_s,0,0) \, , \qquad H, H_s > 0 \, ,
\end{equation}
i.e., magnetic and staggered fields are aligned, and they both point into the direction of the order parameter (staggered magnetization at
$T$=0). In presence of these fields, the magnon dispersion relations exhibit an energy gap and their degeneracy is lifted,\footnote{Note
that in Eq.~(\ref{disprelAFHparallel}) -- much like in subsequent effective field theory expressions -- the spin-wave velocity $v$ is set
to one.}
\begin{eqnarray}
\label{disprelAFHparallel}
\omega_{+} & = & \sqrt{{\vec k \,}^2 + \frac{M_s H_s}{\rho_s}} + H \, , \nonumber \\
\omega_{-} & = & \sqrt{{\vec k \,}^2 + \frac{M_s H_s}{\rho_s}} - H \, .
\end{eqnarray}
The leading-order effective constants that appear in the dispersion relations are $M_s$ (staggered magnetization at $T$=0 and infinite
volume) as well as $\rho_s$ (spin stiffness).

An important point to emphasize is that the lower spin-wave excitation $\omega_{-}$ acquires negative values, unless the stability
criterion
\begin{equation}
\label{stabilityCondition}
H_s > \frac{\rho_s}{M_s} \, H^2
\end{equation}
is met. Throughout the present investigation we assume this is the case. On the other hand, if the condition is not satisfied, the
staggered magnetization vector changes its orientation and a new configuration orthogonal to the external magnetic field is realized -- see
Refs.~\citep{Hof17,Hof20b} for an effective field theory analysis of this alternative situation.

The following discussion of entropy and spin order relies on the representation of the free energy density derived in Ref.~\citep{Hof20a}.
The interested reader is invited to consult this reference for technical details regarding the evaluation of the partition function. Here
we merely quote the final renormalized two-loop free energy density of antiferromagnetic films subjected to mutually parallel staggered and
magnetic fields [as defined by Eq.~(\ref{externalFields})],
\begin{equation}
\label{freeEDtwoLoopParallel}
z = z_0 - {\hat g}_0 + \frac{H}{\rho_s} \, {\hat g}_1 \, \frac{\partial {\hat g}_0}{\partial H}
- \frac{\sqrt{M_s H_s} H}{4 \pi \rho_s^{3/2}} \, \frac{\partial {\hat g}_0}{\partial H}
- \frac{H^2}{\rho_s}{( {\hat g}_1)}^2
+ \frac{\sqrt{M_s H_s} H^2}{2 \pi \rho_s^{3/2}} \, {\hat g}_1 \, ,
\end{equation}
where $z_0$ is the vacuum energy density (free energy density at zero temperature),
\begin{equation}
z_0 = - M_s H_s - \frac{M^{3/2}_s H^{3/2}_s}{6 \pi \rho_s^{3/2}} - (k_2 + k_3) \frac{M^2_s H^2_s}{\rho_s^2}
- \frac{M_s H_s  H^2}{16 \pi^2 \rho_s^2} \, .
\end{equation}
The finite-temperature portion of the free energy density involves the dimensionful kinematical functions ${\hat g}_r$ -- or, equivalently,
the dimensionless kinematical functions ${\hat h}_r$ -- that are defined as\footnote{The function ${\hat g}_2$ does not show up in the free
energy density, but it is relevant in the staggered and sublattice magnetizations (see below).}
\begin{eqnarray}
\label{g0g1g2Bose}
{\hat g}_0 \! & = &  \! T^3 \, {\int}_{\!\!\!0}^{\infty} \mbox{d} \lambda \, \lambda^{-5/2} e^{-\lambda m^2/4 \pi t^2}
\Bigg\{ \sqrt{\lambda} \, \theta_3\Big( \frac{m_H \lambda}{2 t}, e^{- \pi \lambda} \Big) e^{m_H^2 \lambda/4 \pi t^2} - 1 \Bigg\}
\equiv T^3 \, {\hat h}_0\, , \nonumber \\
{\hat g}_1  \! & = &  \! \frac{T}{4 \pi} \, {\int}_{\!\!\!0}^{\infty} \mbox{d} \lambda \, \lambda^{-3/2} e^{-\lambda m^2/4 \pi t^2}
\Bigg\{ \sqrt{\lambda} \, \theta_3\Big( \frac{m_H \lambda}{2 t}, e^{- \pi \lambda} \Big) e^{m_H^2 \lambda/4 \pi t^2} - 1 \Bigg\}
\equiv T \, {\hat h}_1 \, , \\
{\hat g}_2 \!  & = &  \! \frac{1}{16 \pi^2 T} \, {\int}_{\!\!\!0}^{\infty} \mbox{d} \lambda \, \lambda^{-1/2} e^{-\lambda m^2/4 \pi t^2}
\Bigg\{ \sqrt{\lambda} \, \theta_3\Big( \frac{m_H \lambda}{2t}, e^{- \pi \lambda} \Big)e^{m_H^2 \lambda/4 \pi t^2} - 1 \Bigg\}
\equiv \frac{{\hat h}_2}{T}  \, . \nonumber
\end{eqnarray}
These quantities involve the Jacobi theta function
\begin{equation}
\label{Jacobi3}
\theta_3(u,q) = 1 + 2 \sum_{n=1}^{\infty} q^{n^2} \cos(2 n u) \, ,
\end{equation}
and depend on three dimensionless parameters,
\begin{equation}
\label{definitionRatios}
m \equiv \frac{\sqrt{M_s H_s}}{2 \pi \rho_s^{3/2}} \, , \qquad
m_H \equiv \frac{H}{2 \pi \rho_s} \, \, , \qquad
t \equiv \frac{T}{2 \pi \rho_s} \, .
\end{equation}
Since the common denominator,
\begin{equation}
2 \pi \rho_s \approx J \, ,
\end{equation}
approximately corresponds to the exchange coupling $J$ that sets the microscopic scale, the parameters $m, m_H, t$ ought to be small as the
effective field theory describes the physics in the low-energy sector.

Apart from the leading-order effective constants $M_s$ and $\rho_s$ in the free energy density, Eq.~(\ref{freeEDtwoLoopParallel}), in the
zero-temperature piece $z_0$, two additional effective constants -- $k_2$ and $k_3$ -- show up: so-called next-to-leading order (NLO)
effective constants. All these constants depend on the lattice geometry. Concrete numerical values for $\rho_s$ and $M_s$, as well as for
the spin-wave velocity $v$, are available for the spin-$\frac{1}{2}$ square-lattice antiferromagnet (see Ref.~\citep{GHJNW09}),
\begin{equation}
\label{squareLEC}
\rho_s = 0.1808(4) J \, , \quad M_s = 0.30743(1) / a^2 \, , \quad v = 1.6585(10) J a \, ,
\end{equation}
and for the  spin-$\frac{1}{2}$ honeycomb-lattice antiferromagnet (see Ref.~\citep{JKNW08}),
\begin{equation}
\label{honeyLEC}
\rho_s = 0.102(2) J \, , \quad {\tilde M_s} = 0.2688(3) \, , \quad v = 1.297(16) J a \, ,
\end{equation}
with
\begin{equation}
{\tilde M_s} = \frac{ 3 \sqrt{3}}{4} \, M_s \, a^2 \, .
\end{equation}
Furthermore, for the spin-$\frac{1}{2}$ square-lattice antiferromagnet, the relevant combination $k_2 + k_3$ of next-to-leading order
effective constants that mattes at zero temperature, has been determined in Ref.~\citep{GHJNW09} as
\begin{equation}
\label{k2k3}
\frac{k_2 + k_3}{v^2} = \frac{-0.0037}{2 \rho_s} = \frac{-0.0102}{J} \, .
\end{equation}

It should be emphasized that next-to-leading order effective constants ($k_2, k_3$) only show up in the (zero-temperature) vacuum energy
density $z_0$. The finite-temperature physics of the system, up to two-loop order, is thus fully described in terms of the leading-order
effective constants $\rho_s$ and $M_s$. The only difference between, e.g., square- and honeycomb-lattice antiferromagnets consists in the
concrete numerical values of $\rho_s$ and $M_s$.

The effective field theory is tied to the low-energy sector where the parameters $m, m_H, t$ -- defined in Eq.~(\ref{definitionRatios}) --
are small. In the present investigation we consider values up to
\begin{equation}
\label{domain}
m, m_H, t \ \lesssim 0.4 \, .
\end{equation}
A further constraint is imposed by the stability criterion, Eq.~(\ref{stabilityCondition}), which we implement by choosing the parameter
region of external fields as
\begin{equation}
m > m_H + \delta \, , \qquad  \delta = 0.1 \, .
\end{equation}

\subsection{Entropy Density}
\label{entropy}

In this subsection we derive the low-temperature representation for the two-loop entropy density and study how it is affected by the
external fields. The entropy density $s$ of antiferromagnetic films subjected to mutually parallel staggered and magnetic fields is readily
obtained from the finite-temperature part of the free energy density, $z-z_0$, via
\begin{equation}
s = \frac{\mbox{d}}{\mbox{d} T} \, \Big( z_0 -z \Big) \, ,
\end{equation}
with the result
\begin{eqnarray}
s(t,m,m_H) & = & s_1 T^2 + s_2 T^3 + {\cal O}(T^4) \, .
\end{eqnarray}
The respective coefficients are
\begin{eqnarray}
s_1 & = & \frac{t^2}{2} \, \frac{{\mbox{d}}^2 {\hat h}_{-1}}{\mbox{d} m_H^2}
+ \frac{m m_H t^2}{4} \, \frac{{\mbox{d}}^3 {\hat h}_{-1}}{\mbox{d} m_H^3}
- m_H (1 + \frac{m}{2}) \, \frac{\mbox{d} {\hat h}_0}{\mbox{d} m_H} \nonumber \\
& & - m m_H^2 \frac{{\mbox{d}}^2 {\hat h}_0}{\mbox{d} m_H^2}
+ \frac{m m_H^3}{t^2} \, \frac{\mbox{d} {\hat h}_1}{\mbox{d} m_H} \, , \nonumber \\
s_2 & = & - \frac{m_H t^2}{2 \rho_s} \, \frac{{\mbox{d}}^2 {\hat h}_0}{\mbox{d} m_H^2} \, \frac{\mbox{d} {\hat h}_0}{\mbox{d} m_H}
+ \frac{m^2_H}{\rho_s} \, \frac{\mbox{d} {\hat h}_0}{\mbox{d} m_H} \, \frac{\mbox{d} {\hat h}_1}{\mbox{d} m_H}
- \frac{m_H t^2}{2 \rho_s} \, {\hat h}_1 \, \frac{{\mbox{d}}^3 {\hat h}_{-1}}{\mbox{d} m_H^3} \nonumber \\
& & + \frac{2 m^2_H}{\rho_s} \, {\hat h}_1 \, \frac{{\mbox{d}}^2 {\hat h}_0}{\mbox{d} m_H^2}
- \frac{2 m^3_H}{t^2 \rho_s} \, {\hat h}_1 \, \frac{\mbox{d} {\hat h}_1}{\mbox{d} m_H}
+ \frac{m_H}{\rho_s} \, {\hat h}_1 \, \frac{\mbox{d} {\hat h}_0}{\mbox{d} m_H} \, .
\end{eqnarray}
The kinematical function ${\hat h}_{-1}$ reads
\begin{equation}
\label{gMinus1Bose}
{\hat h}_{-1} = \frac{{\hat g}_{-1}}{T^5} = 4 \pi \, {\int}_{\!\!\!0}^{\infty} \mbox{d} \lambda \, \lambda^{-7/2} e^{-\lambda m^2/4 \pi t^2}
\Bigg\{ \sqrt{\lambda} \, \theta_3\Big( \frac{m_H \lambda}{2 t}, e^{- \pi \lambda} \Big) e^{m_H^2 \lambda/4 \pi t^2} - 1 \Bigg\} \, .
\end{equation}
In the derivation of the above representation for the entropy density we have used the relation
\begin{equation}
\frac{\mbox{d} {\hat g}_r}{\mbox{d} T} = \frac{1}{2T} \, \frac{{\mbox{d}}^2 {\hat g}_{r-1}}{\mbox{d} H^2}
- \frac{H}{T} \, \frac{\mbox{d} {\hat g}_r}{\mbox{d} H}  \, ,
\end{equation}
which is based on the identities
\begin{eqnarray}
e^{-\frac{n^2 \beta^2}{4 \lambda}} n \, e^{\mp n \beta H} & = & \mp \frac{1}{\beta} \, \frac{\partial}{\partial H} \Bigg\{ e^{-\frac{n^2 \beta^2}{4 \lambda}}
e^{\mp n \beta H}  \Bigg\} \, , \nonumber \\
e^{-\frac{n^2 \beta^2}{4 \lambda}} n^2 \, e^{\mp n \beta H} & = & \frac{1}{\beta^2} \, \frac{\partial^2}{\partial H^2}
\Bigg\{ e^{-\frac{n^2 \beta^2}{4 \lambda}} e^{\mp n \beta H}  \Bigg\} \, .
\end{eqnarray}
Recall that the kinematical functions ${\hat g}_0, {\hat g}_1, {\hat g}_2$ are defined in Eq.~(\ref{g0g1g2Bose}).

\begin{figure}
\begin{center}
\hbox{
\includegraphics[width=7.5cm]{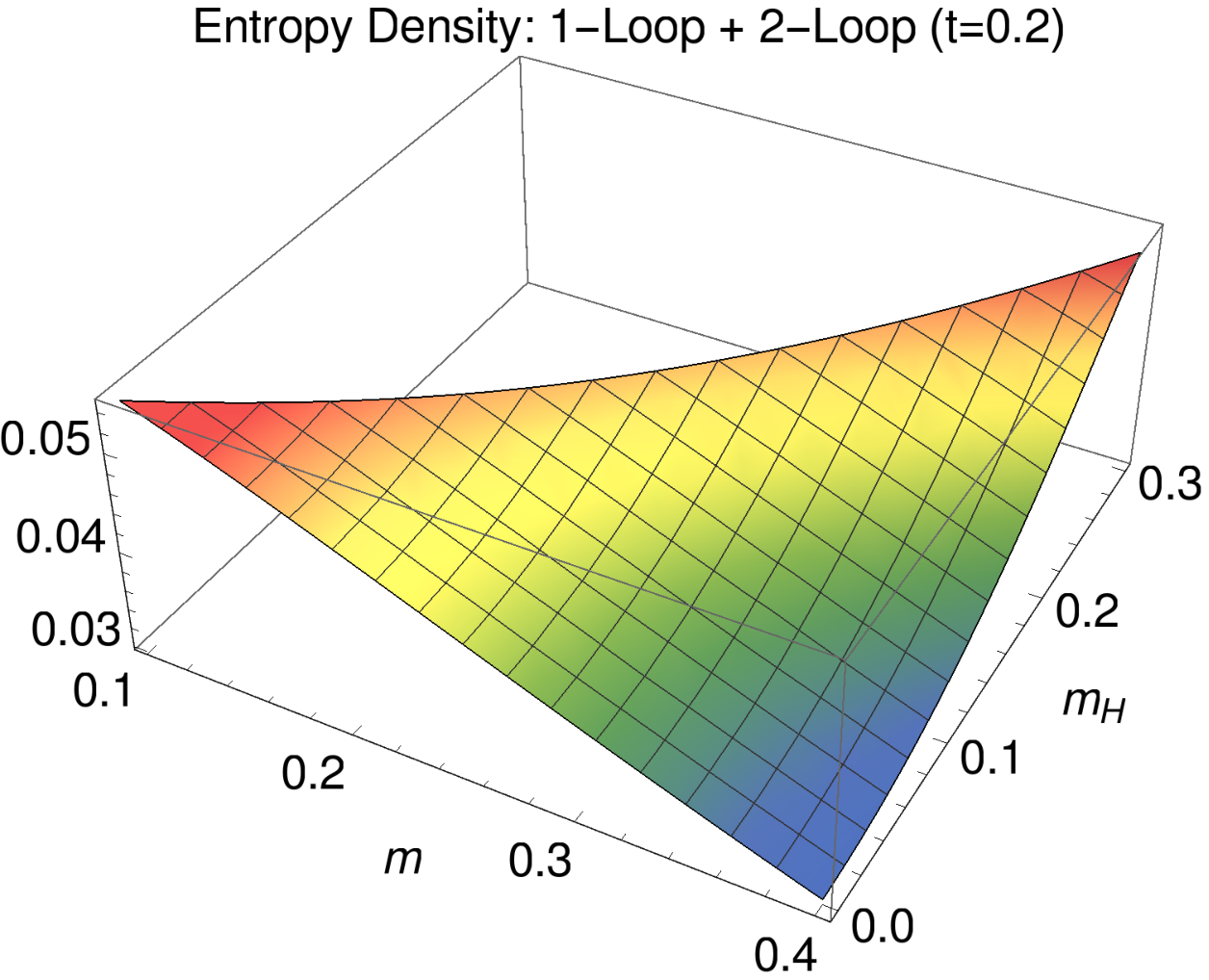}
\hspace{2mm}
\includegraphics[width=7.5cm]{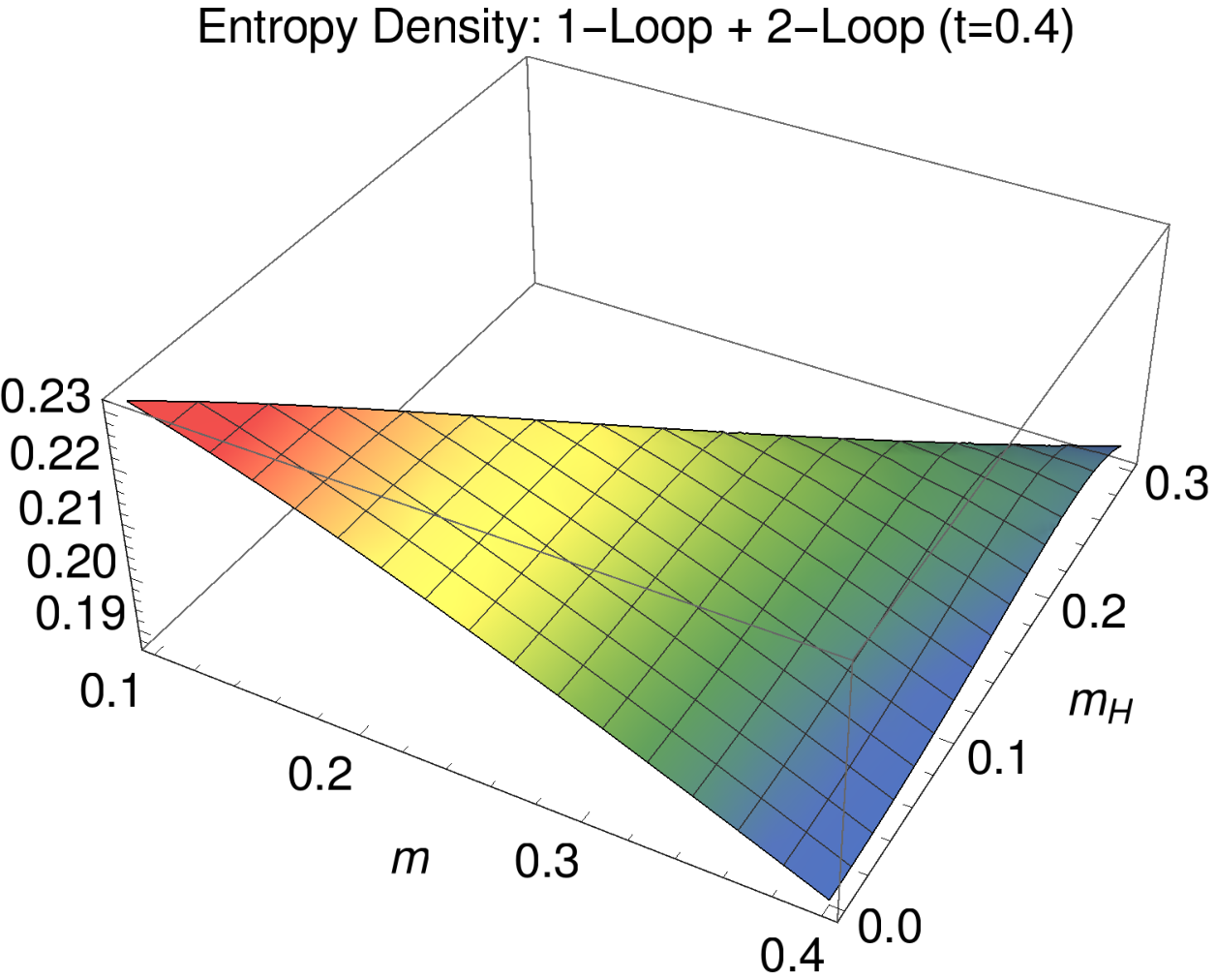}}
\end{center}
\caption{[Color online] Entropy density $s$ for the spin-$\frac{1}{2}$ square-lattice antiferromagnet in mutually parallel magnetic ($m_H$)
and staggered ($m$) fields at the temperatures $t=0.2$ and $t=0.4$.}
\label{figure1}
\end{figure}

In Fig.~\ref{figure1}, for the temperatures $t = \{0.2,0.4\}$, we depict the entropy density, i.e., the quantity
\begin{equation}
s_1 T^2 + s_2 T^3 \, ,
\end{equation}
as a function of magnetic ($m_H$) and staggered ($m$) field strength. Roughly, two opposite tendencies can be identified: (1) If the
staggered field becomes stronger, the entropy decreases, and (2) if the magnetic field becomes stronger, the entropy increases. Naively,
one would conclude that the staggered field establishes spin order by enforcing antialignment of the spins, while the magnetic field
perturbs the antialigned array of spins. However, as the figure referring to the more elevated temperature $t=0.4$ indicates, in stronger
fields subtle effects emerge. To interpret these observations, let us also consider how uniform and staggered magnetization -- as well as
the individual sublattice magnetizations -- behave when magnetic field, staggered field and temperature are varied. 

\subsection{Order-Disorder Effects}
\label{OrderDisorder}

The low-temperature series for the staggered magnetization $M_s$ and the uniform magnetization $M$ have been derived in
Ref.~\citep{Hof20a}. The staggered magnetization takes the form
\begin{equation}
\label{OPAF}
M_s(t,m,m_H) = M_s(0,m,m_H) + {\tilde \sigma}_1 T + {\tilde \sigma}_2 T^2 + {\cal O}(T^3) \, ,
\end{equation}
with coefficients
\begin{eqnarray}
{\tilde \sigma}_1(t,m,m_H) & = & - \frac{M_s}{\rho_s} \, {\hat h}_1 \, , \nonumber \\
{\tilde \sigma}_2(t,m,m_H) & = & \frac{M_s}{\rho_s} \, \Bigg\{ \frac{m_H}{\rho_s} \, {\hat h}_2 \, \frac{\partial{\hat h}_0}{\partial m_H}
+ \frac{m_H}{\rho_s} \, {\hat h}_1 \, \frac{\partial{\hat h}_1}{\partial m_H}
+ \frac{m_H t}{8 \pi \rho_s m} \, \frac{\partial{\hat h}_0}{\partial m_H}  \nonumber \\
& & - \frac{m m_H}{4\pi \rho_s t} \, \frac{\partial{\hat h}_1}{\partial m_H}
- \frac{2 m_H^2}{\rho_s t^2} \, {\hat h}_1 {\hat h}_2
- \frac{m_H^2}{4\pi \rho_s m t} \, {\hat h}_1
+ \frac{m m_H^2}{2 \pi \rho_s t^3} \, {\hat h}_2 \Bigg\} \, ,
\end{eqnarray}
and the uniform magnetization amounts to
\begin{equation}
\label{magnetizationAF}
M(t,m,m_H) = M(0,m,m_H) + {\hat \sigma}_1 T + {\hat \sigma}_2 T^2 + {\cal O}(T^3) \, ,
\end{equation}
with coefficients
\begin{eqnarray}
{\hat \sigma}_1(t,m,m_H) & = & 2 \pi \rho_s t^2 \frac{\partial {\hat h}_0}{\partial m_H} \, , \nonumber \\
{\hat \sigma}_2(t,m,m_H) & = & - 2 \pi t^2 {\hat h}_1 \frac{\partial {\hat h}_0}{\partial m_H}
- 2 \pi m_H t^2 \frac{\partial {\hat h}_1}{\partial m_H} \frac{\partial {\hat h}_0}{\partial m_H}
- 2 \pi m_H t^2 {\hat h}_1 \frac{\partial^2 {\hat h}_0}{\partial m^2_H} \nonumber \\
& & + \frac{m t}{2} \frac{\partial {\hat h}_0}{\partial m_H}
+ \frac{m m_H t}{2} \frac{\partial^2 {\hat h}_0}{\partial m^2_H}
+ 4 \pi m_H {({\hat h}_1)}^2
+ 4 \pi m^2_H {\hat h}_1 \frac{\partial {\hat h}_1}{\partial m_H} \nonumber \\
& & - \frac{2 m m_H}{t} {\hat h}_1
- \frac{m m_H^2}{t} \frac{\partial {\hat h}_1}{\partial m_H} \, .
\end{eqnarray}
Although the discussion below mainly concerns effects caused by finite temperature, for completeness and clarity we also quote the results
for the staggered and uniform magnetization at zero temperature. These are
\begin{equation}
\label{OPT0}
\frac{M_s(0,m,m_H)}{M_s} = 1 + \frac{m}{2} + \frac{m_H^2}{4} + 8 \pi^2 \rho_s (k_2 + k_3) m^2 \, ,
\end{equation}
and
\begin{equation}
\label{MagT0}
M(0,m,m_H) = \pi \rho^2_s m^2 m_H \, ,
\end{equation}
respectively. The numerical value of the combination $k_2 + k_3$ of next-to-leading order effective constants for the spin-$\frac{1}{2}$
square-lattice antiferromagnet is given in Eq.~(\ref{k2k3}). While our effective results apply to any bipartite lattice, in the plots that
follow, we specifically refer to square-lattice geometry and spin-$\frac{1}{2}$, since only for this system all relevant effective
constants are known.

\begin{figure}
\begin{center}
\hbox{
\includegraphics[width=7.5cm]{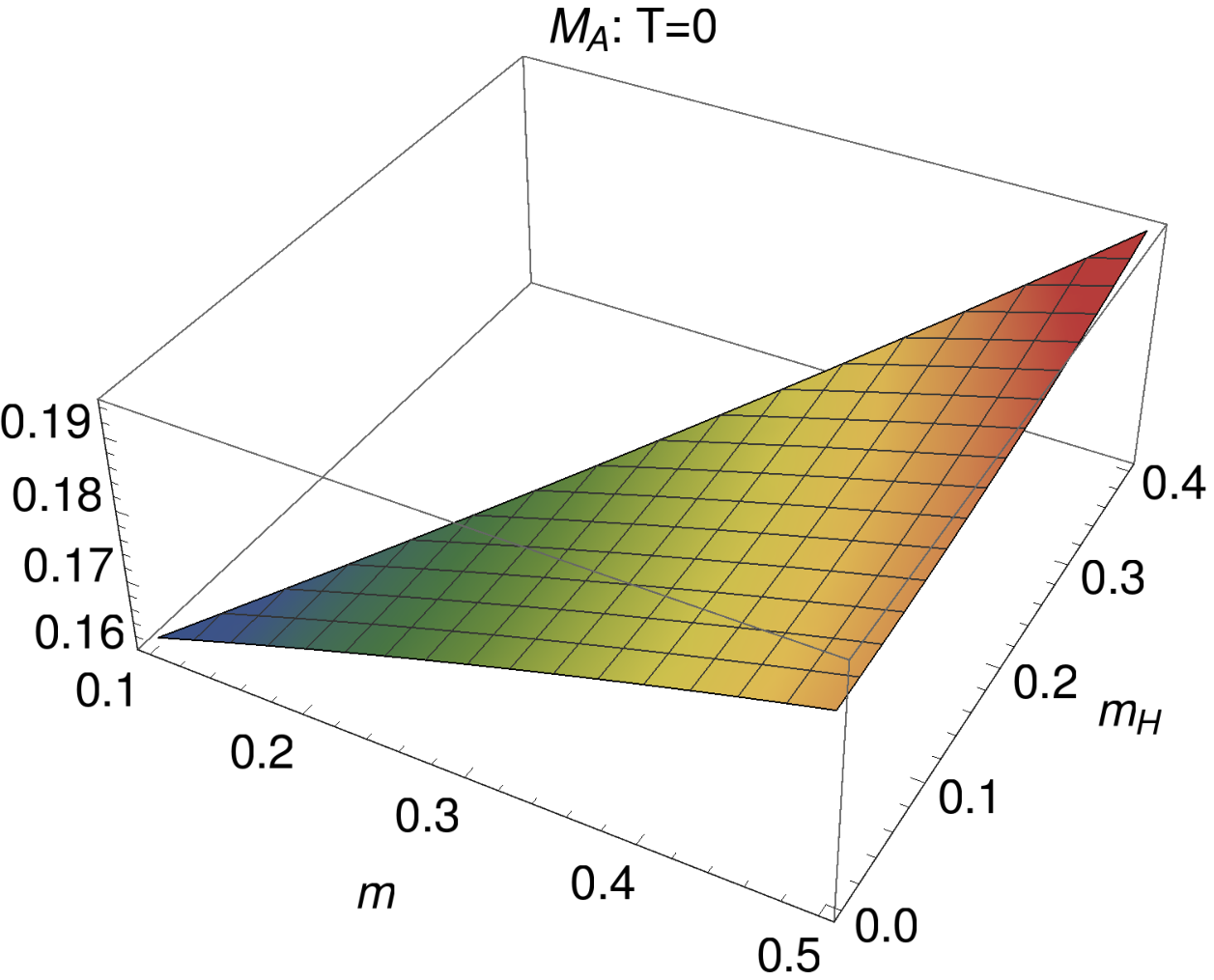}
\hspace{2mm}
\includegraphics[width=7.5cm]{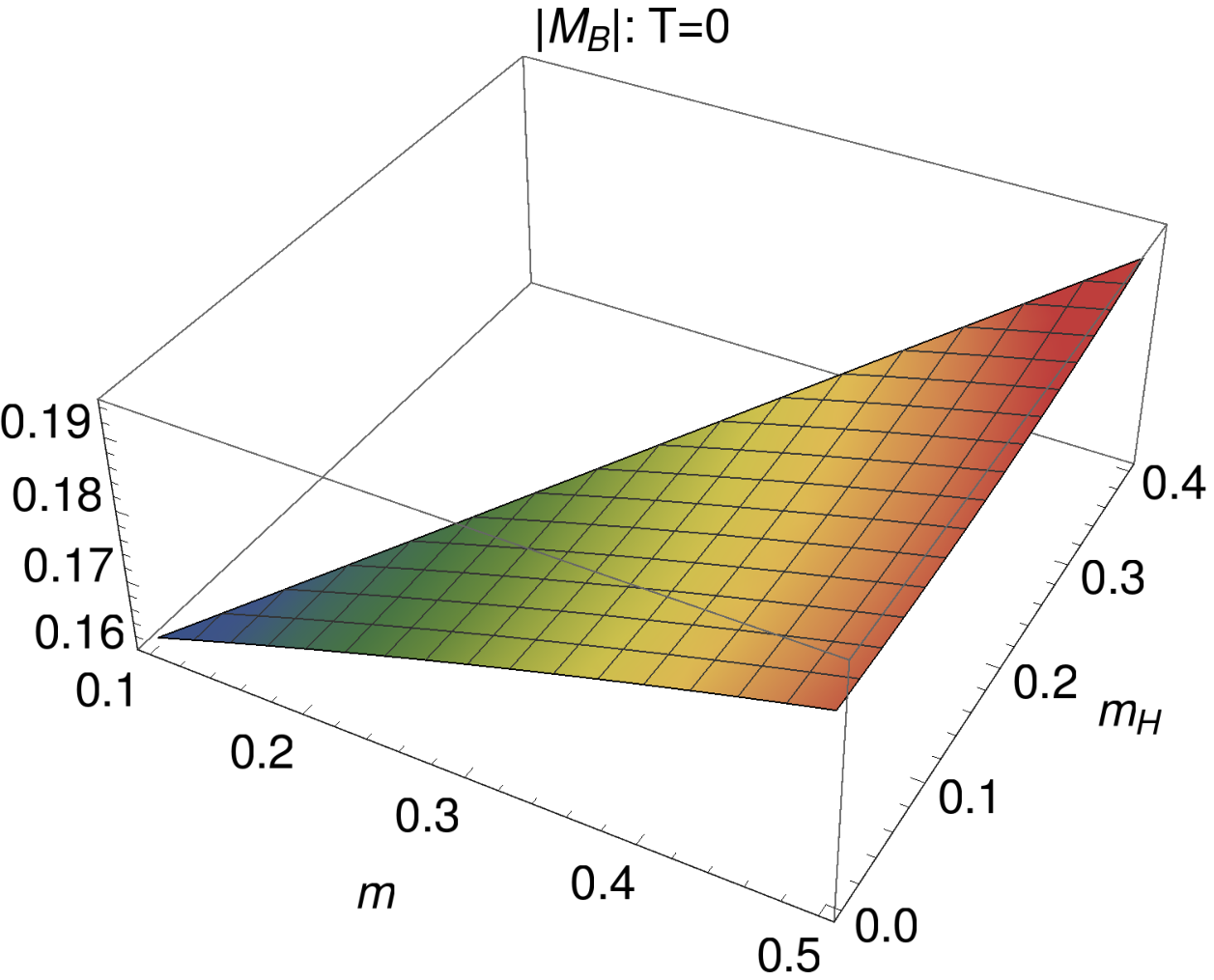}}
\end{center}
\caption{[Color online] Sublattice magnetizations $M_A$ and $|M_B|$ for the spin-$\frac{1}{2}$ square-lattice antiferromagnet in mutually
parallel magnetic ($m_H$) and staggered ($m$) fields at $T$=0.}
\label{figure2}
\end{figure}

Apart from $M_s$ and $M$, it is illuminating to also consider the individual sublattice magnetizations $M_A$ and $M_B$. On a bipartite
lattice they can be extracted from the uniform and staggered magnetization as
\begin{eqnarray}
\label{sublatticeMag}
M_A(t,m,m_H) = \frac{M(t,m,m_H) + M_s(t,m,m_H)}{2} \, , \nonumber \\
M_B(t,m,m_H) = \frac{M(t,m,m_H) - M_s(t,m,m_H)}{2} \, .
\end{eqnarray}
By definition, $M_A$ is positive ("A-spins point up") and $M_B$ is negative ("B-spins point down"). While figures for the zero-temperature
uniform and staggered magnetizations in external fields have been presented in Ref.~\citep{Hof20a} (Figs.~3 and 4, respectively), here, in
Fig.~\ref{figure2}, we depict the zero-temperature sublattice magnetizations $M_A(0,m,m_H)$ and $|M_B(0,m,m_H)|$. One observes that both
$M_A$ and $|M_B|$ increase when the magnetic or the staggered field become stronger. The explanation is via suppression of quantum
fluctuations caused by the external fields. Note that the magnitude of the net external field on sublattice $A$ is not the same as on
sublattice $B$. The net external field on sublattice $A$ is given by the sum $H_s+H$, whereas on sublattice $B$ it is given by the
difference $H_s-H$. Due to the stronger net external field on sublattice $A$, the suppression of quantum fluctuations on sublattice $A$ is
more pronounced. Accordingly, the zero-temperature sublattice magnetization $M_A$ is larger than the zero-temperature sublattice
magnetization $|M_B|$ for any values of fixed external fields.

\begin{figure}
\begin{center}
\hbox{
\includegraphics[width=7.3cm]{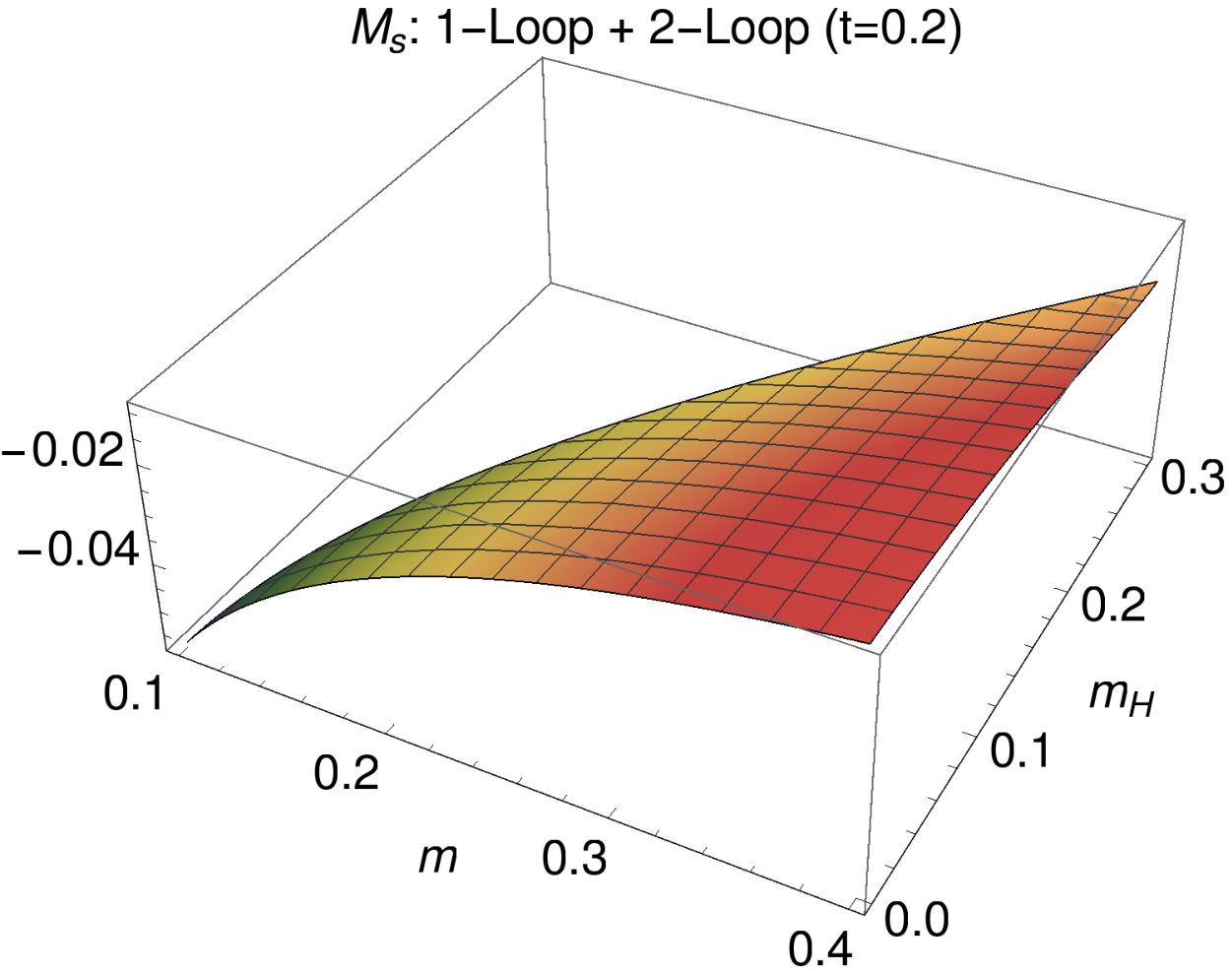} 
\hspace{2mm}
\includegraphics[width=7.3cm]{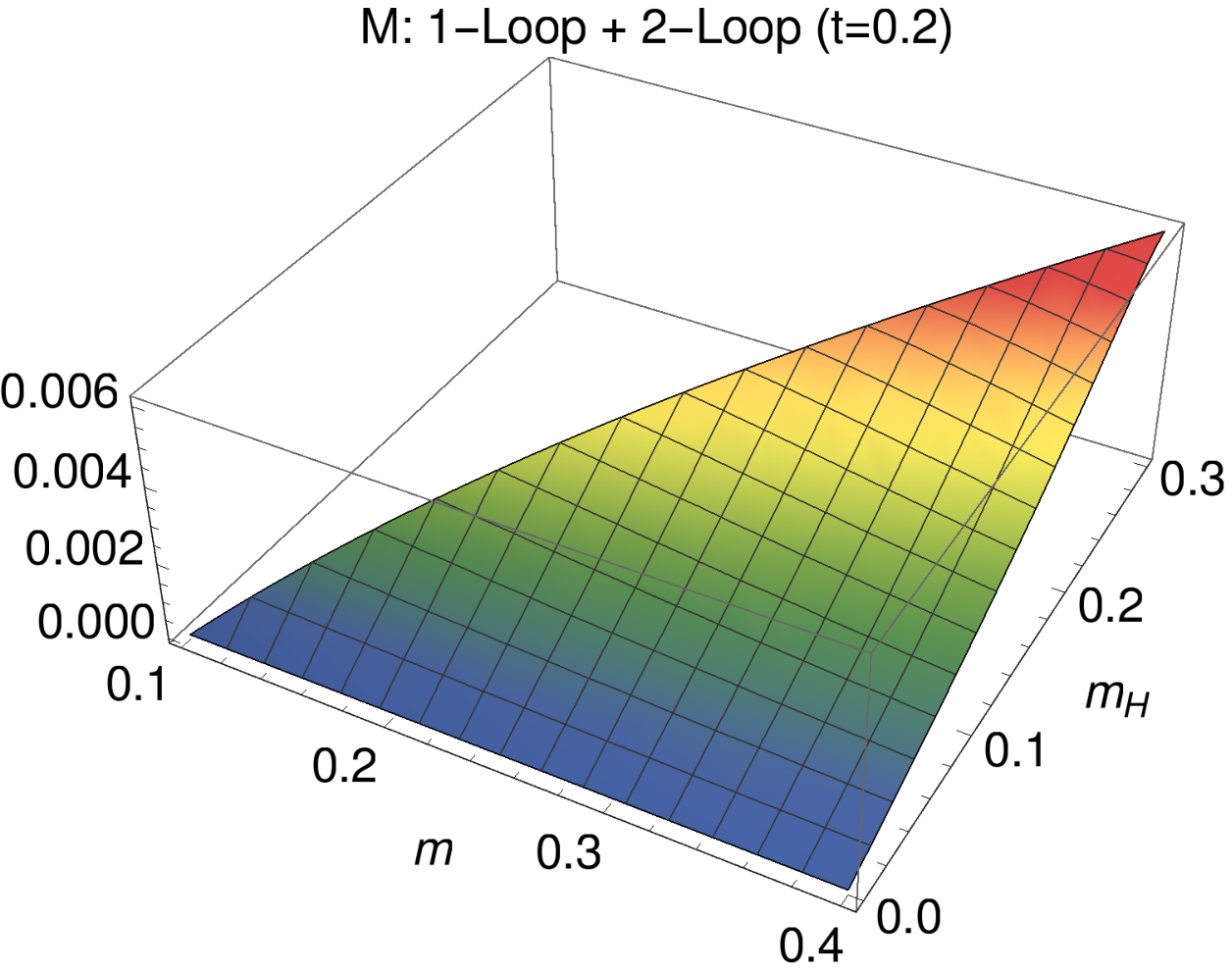}}
\vspace{2mm}
\hbox{
\includegraphics[width=7.3cm]{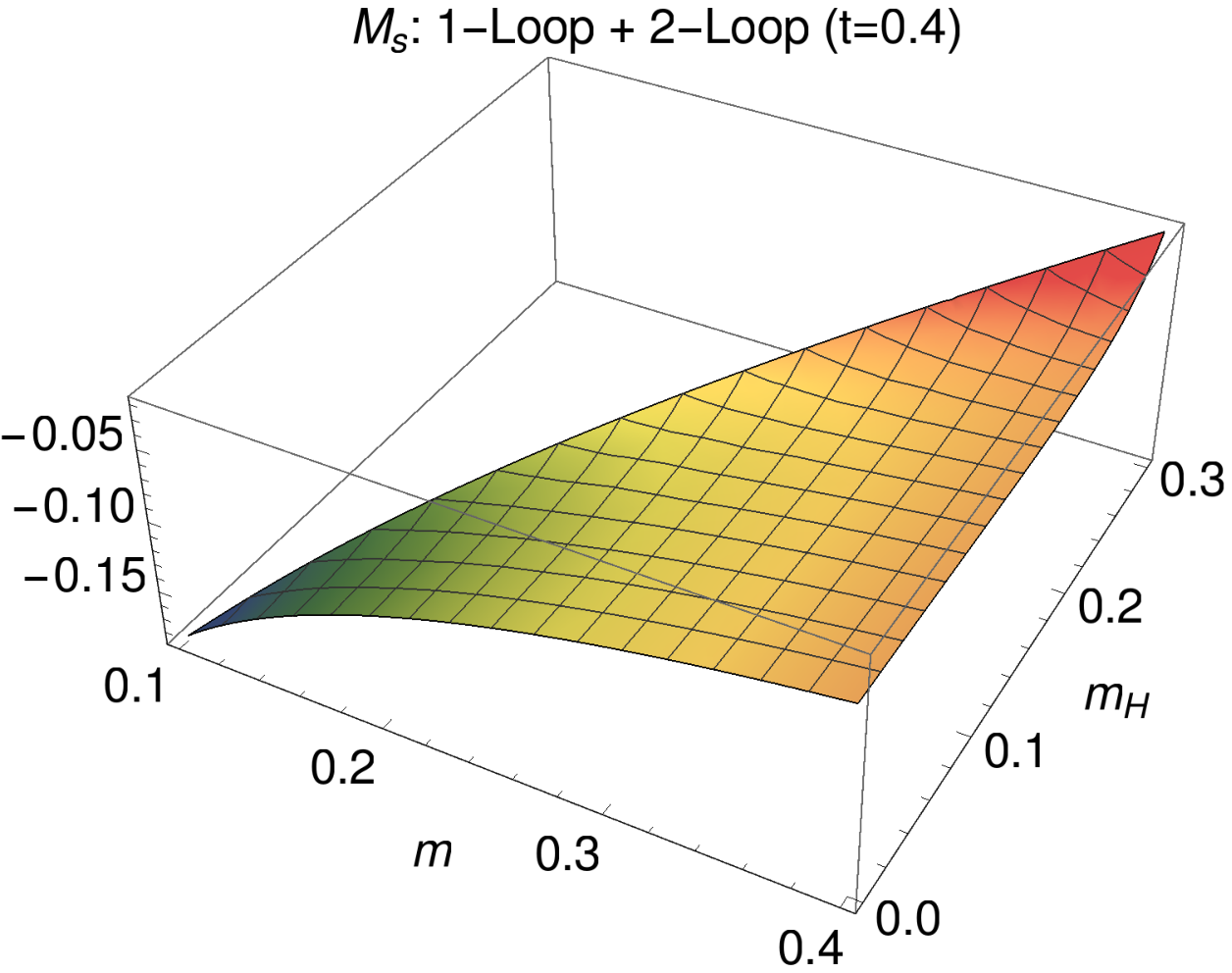} 
\hspace{2mm}
\includegraphics[width=7.3cm]{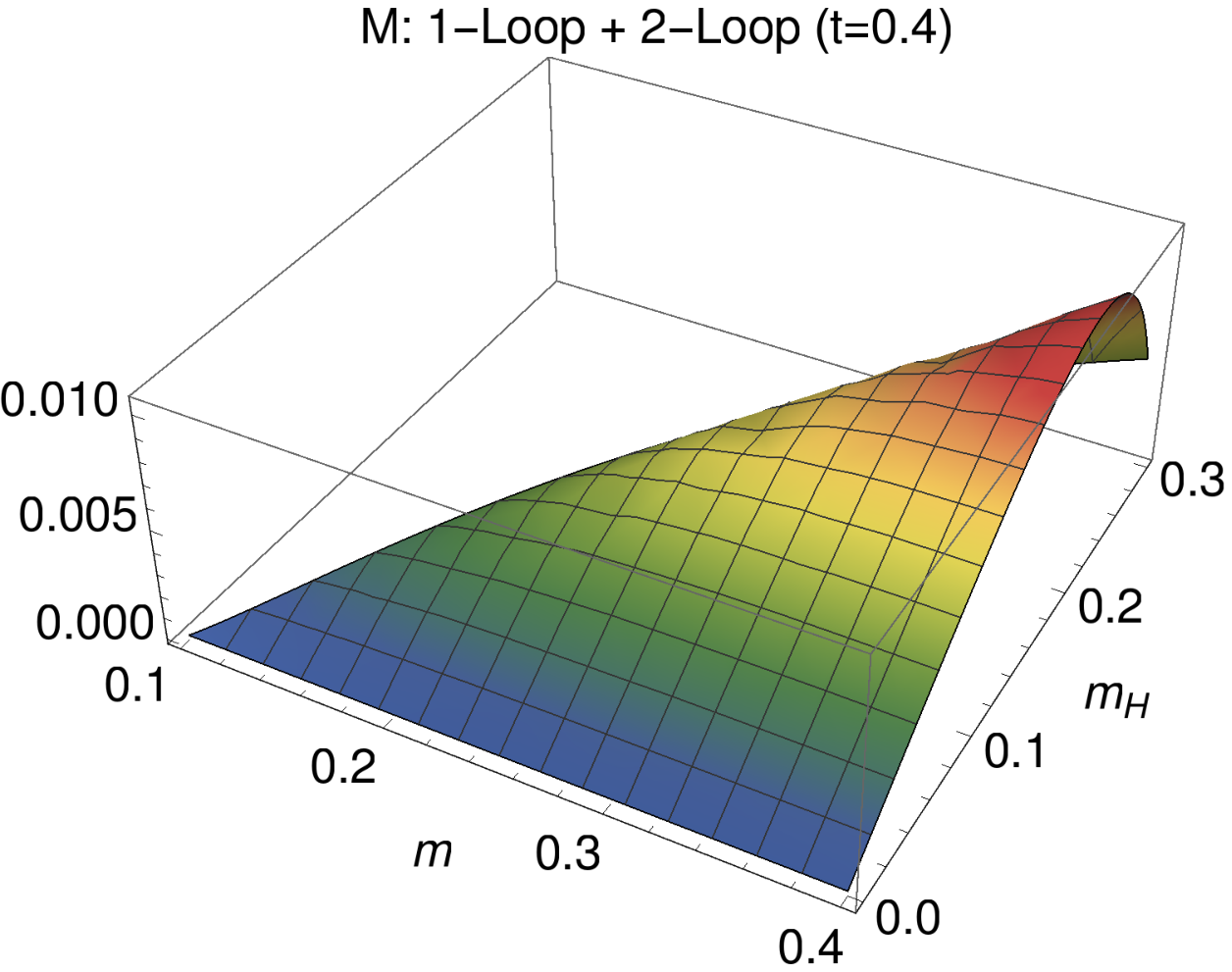}}
\end{center}
\caption{[Color online] Staggered and uniform magnetization for the spin-$\frac{1}{2}$ square-lattice antiferromagnet at the temperatures
$t = 0.2$ (upper panel) and $t = 0.4$ (lower panel): Dependence on  mutually parallel magnetic ($m_H$) and staggered ($m$) fields.}
\label{figure3}
\end{figure}

We now turn to finite temperature where rather subtle effects emerge. In Fig.~\ref{figure3} we first show staggered and uniform
magnetization as a function of $m_H$ and $m$ for the temperatures $t=0.2$ and $t=0.4$, as in the previous plots for the entropy density.
More precisely, we depict the sum of 1-loop and 2-loop contributions given in Eqs.~(\ref{OPAF}) and (\ref{magnetizationAF}),
\begin{eqnarray}
& M_s: & {\tilde \sigma}_1 T + {\tilde \sigma}_2 T^2 \, , \nonumber \\
& M: & {\hat \sigma}_1 T + {\hat \sigma}_2 T^2 \, ,
\end{eqnarray}
without superimposing the dominant $T$=0 contributions. The plots in Fig.~\ref{figure3} hence capture the change of staggered and uniform
magnetizations when temperature is raised from $t$=0 to $t=\{0.2, 0.4\}$. These are the relevant quantities to be compared with entropy
density where the zero-temperature piece is excluded as well.\footnote{Notice that uniform and sublattice magnetizations in the plots are
given in the same units as the staggered magnetization: $1/a^2$.} As expected, the change in the staggered magnetization is negative, i.e.,
the staggered magnetization drops when temperature is raised -- while magnetic and staggered field strengths held fixed -- due to the
thermal fluctuations. The decrease is most pronounced in weak fields where the thermal disruption of spin order is strongest.

Interestingly, in the uniform magnetization we observe a qualitatively different pattern: if temperature is raised from $t$=0 to
$t=\{0.2, 0.4\}$ -- while magnetic and staggered fields held fixed -- the change in the uniform magnetization is positive. Here the
combined effect of thermal fluctuations and suppression of quantum fluctuations by the external fields is rather counterintuitive: the
uniform magnetization is enhanced -- and not weakened. The enhancement becomes larger as the magnetic field strength increases. However,
the enhancement does not grow monotonously: at more elevated temperatures ($t=0.4$) -- and in stronger external fields -- the enhancement
becomes less distinctive.

These findings regarding the staggered and the uniform magnetization are consistent with the previous observation (see Fig.~\ref{figure1})
that the entropy density drops when the staggered field becomes stronger: the antialigned spin pattern is enforced. Likewise, they are
consistent with the observation that the entropy density grows when the magnetic field becomes stronger: the antialigned spin pattern is
perturbed. Note that the correlation is between entropy density and the staggered (and not the uniform) magnetization -- the effects in the
uniform magnetization are much less pronounced: there is about a one order of magnitude difference between the changes in $M_s$ and $M$ in
the respective plots of Fig.~\ref{figure3}. Also, at the more elevated temperature $t=0.4$ and in stronger staggered fields, the entropy
density only initially grows when $H$ becomes stronger, but then starts to drop (RHS of Fig.~\ref{figure1}) -- much like the decrease in
the staggered magnetization becomes less distinctive when $H$ gets stronger (RHS and lower panel of Fig.~\ref{figure3}).

\begin{figure}
\begin{center}
\hbox{
\includegraphics[width=7.3cm]{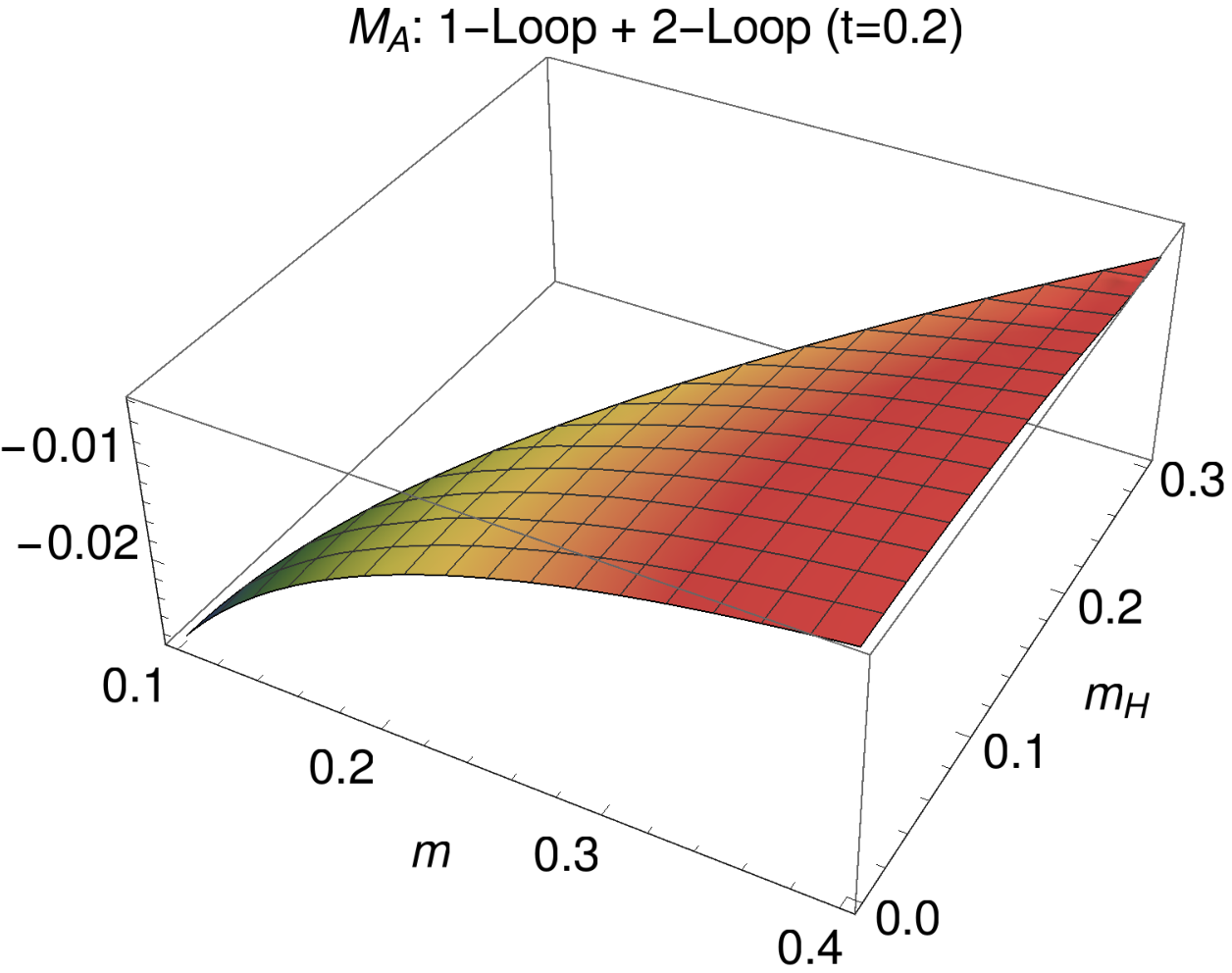} 
\hspace{2mm}
\includegraphics[width=7.3cm]{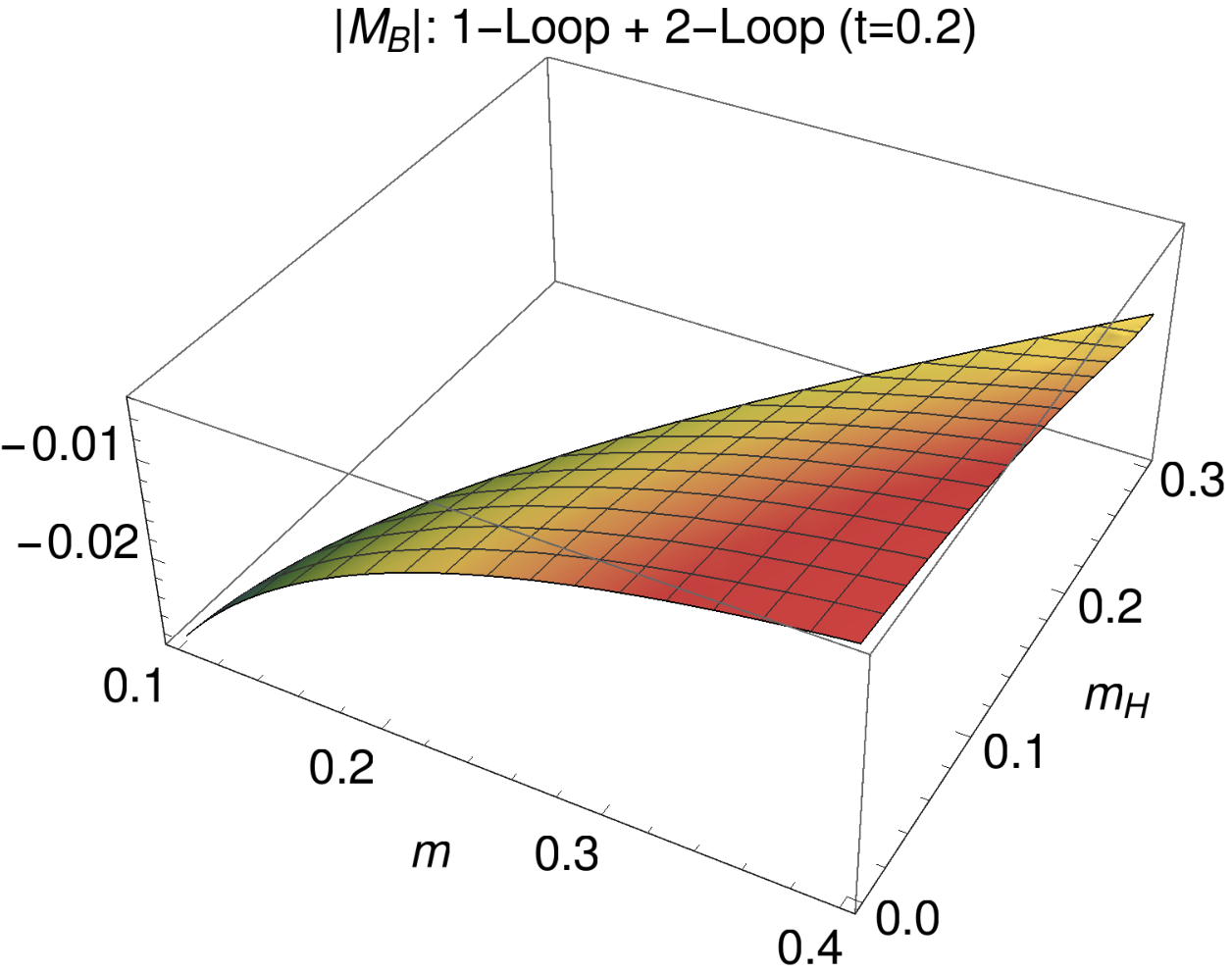}}
\vspace{2mm}
\hbox{
\includegraphics[width=7.3cm]{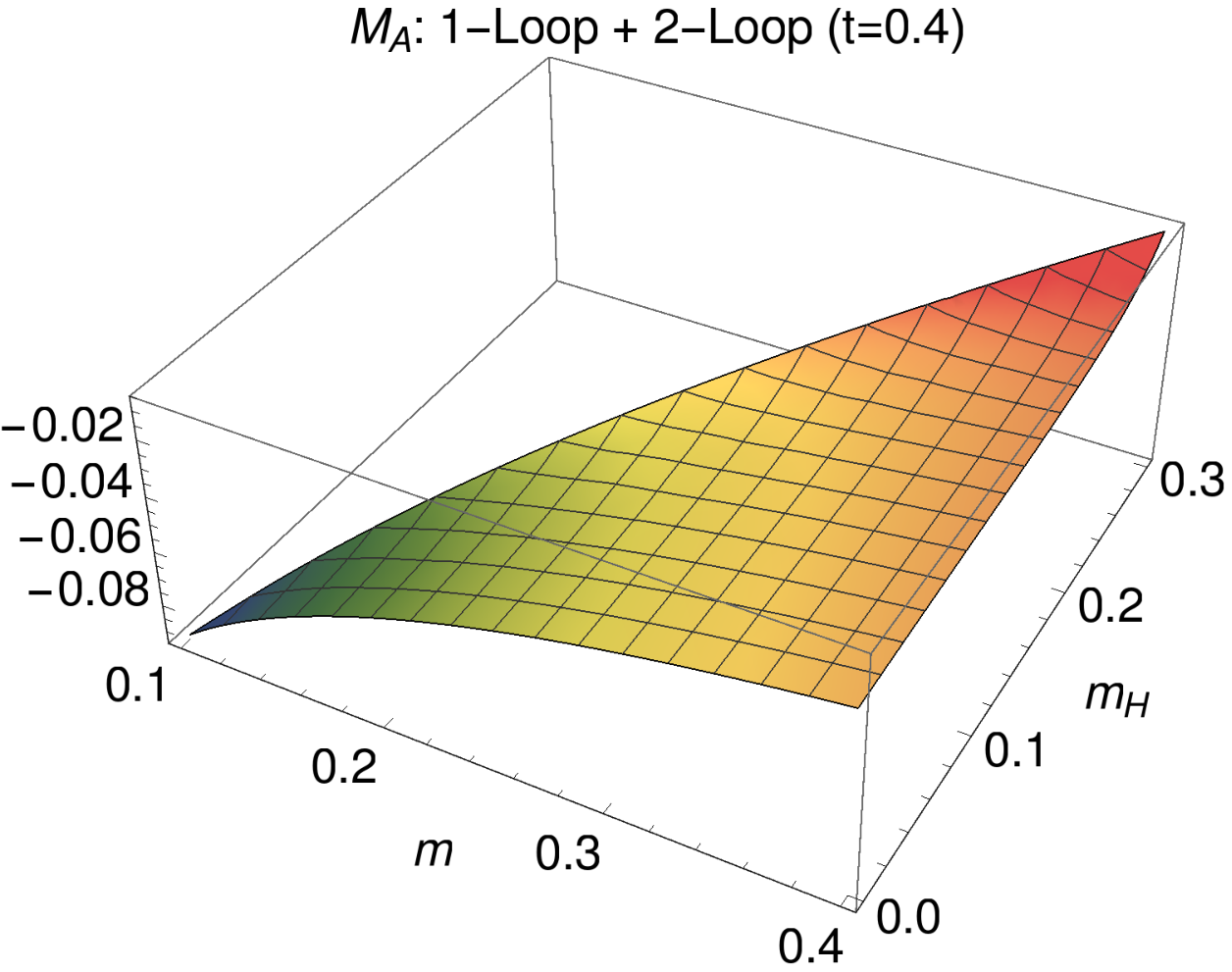} 
\hspace{2mm}
\includegraphics[width=7.3cm]{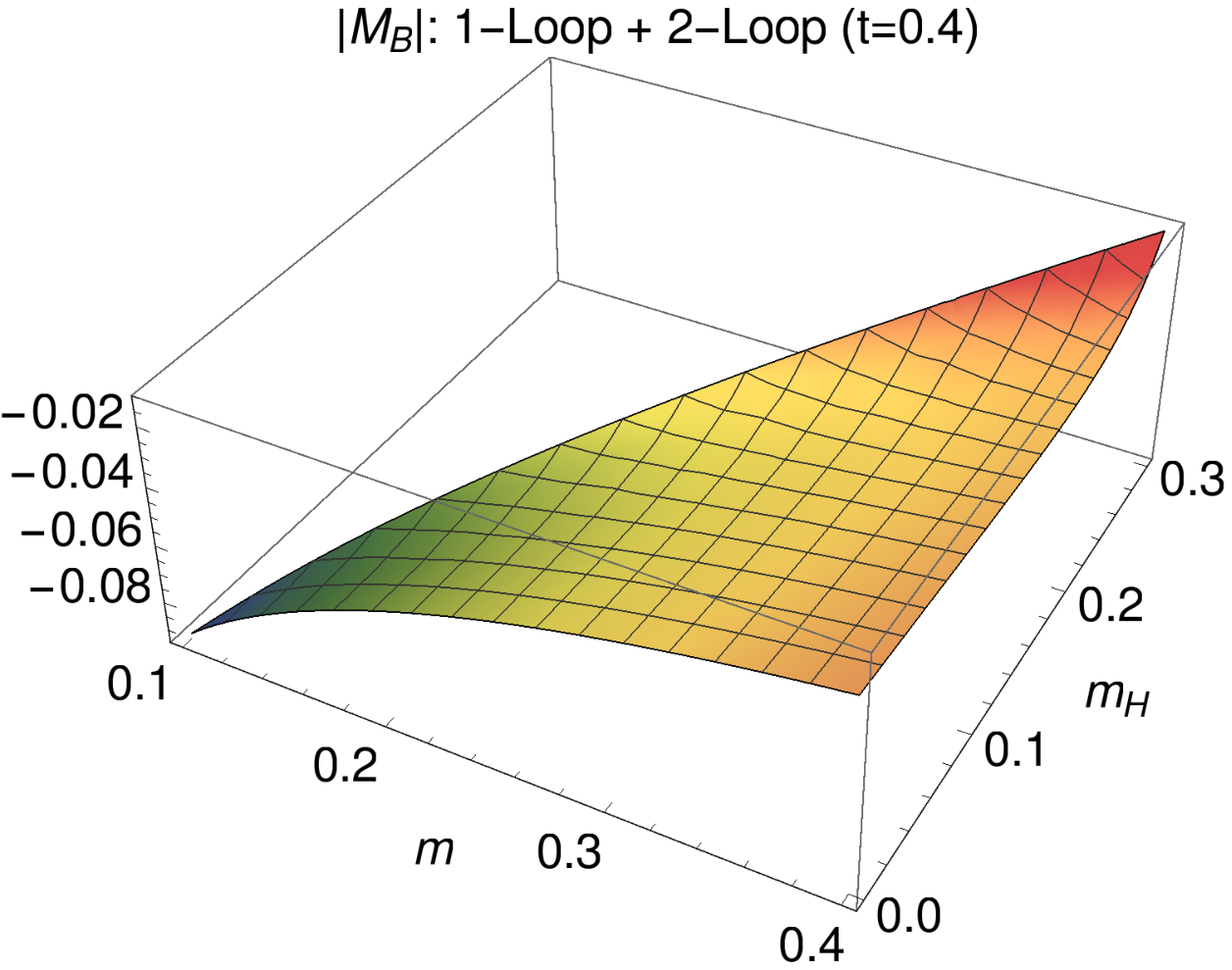}}
\end{center}
\caption{[Color online] Sublattice magnetizations $M_A$ and $|M_B|$ for the spin-$\frac{1}{2}$ square-lattice antiferromagnet at the
temperatures $t = 0.2$ (upper panel) and $t = 0.4$ (lower panel): Dependence on mutually parallel magnetic ($m_H$) and staggered ($m$)
fields.}
\label{figure4}
\end{figure}

This subtle interplay between entropy and spin order can also be appreciated in the behavior of the individual sublattice magnetizations.
As Fig.~\ref{figure4} indicates, the sublattice magnetizations $M_A$ and $|M_B|$ both behave in a qualitatively similar way as the
staggered magnetization: the thermal perturbation of the antialigned spins is most drastic in weak fields. Analogously, at more elevated
temperatures ($t=0.4$) -- and in stronger external fields -- the decrease in both $M_A$ and $|M_B|$ gets less distinctive as the magnetic
field becomes stronger.

So far we have been focusing on just two fixed temperatures. Let us now investigate in more detail how entropy and magnetizations vary with
temperature. To explore this situation, we consider two representative points in parameter space $\{m,m_H\}$ that we choose as
$\{0.3,0.05\}$ and $\{0.3,0.2\}$, respectively. For each point we evaluate entropy density, uniform magnetization, staggered magnetization
and the sublattice magnetizations as a function of temperature. Note that all magnetizations are the magnetizations induced by finite
temperature, i.e., we consider the changes of $M, M_s, M_A, M_B$ when temperature is raised from $t=0$ to $t \neq 0$.

\begin{figure}
\begin{center}
\hbox{
\includegraphics[width=7.5cm]{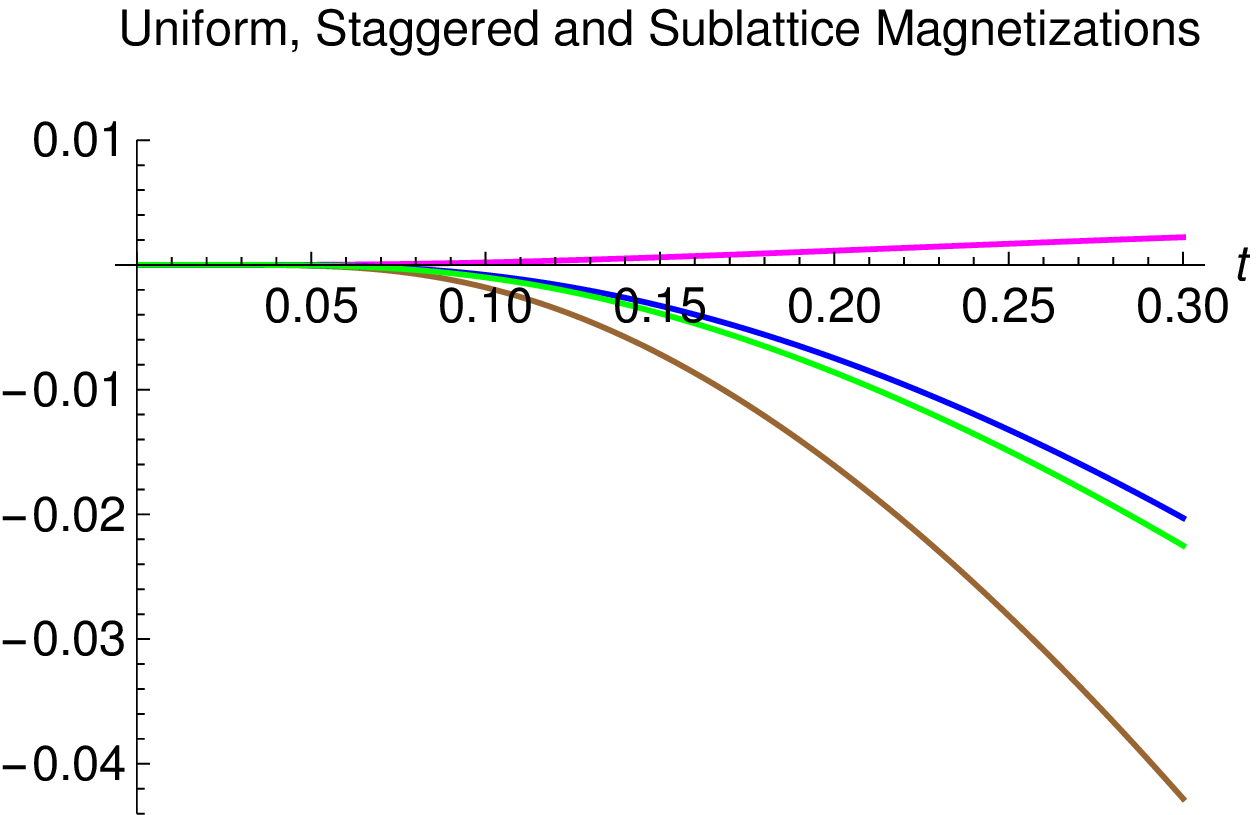}
\hspace{2mm}
\includegraphics[width=6.8cm]{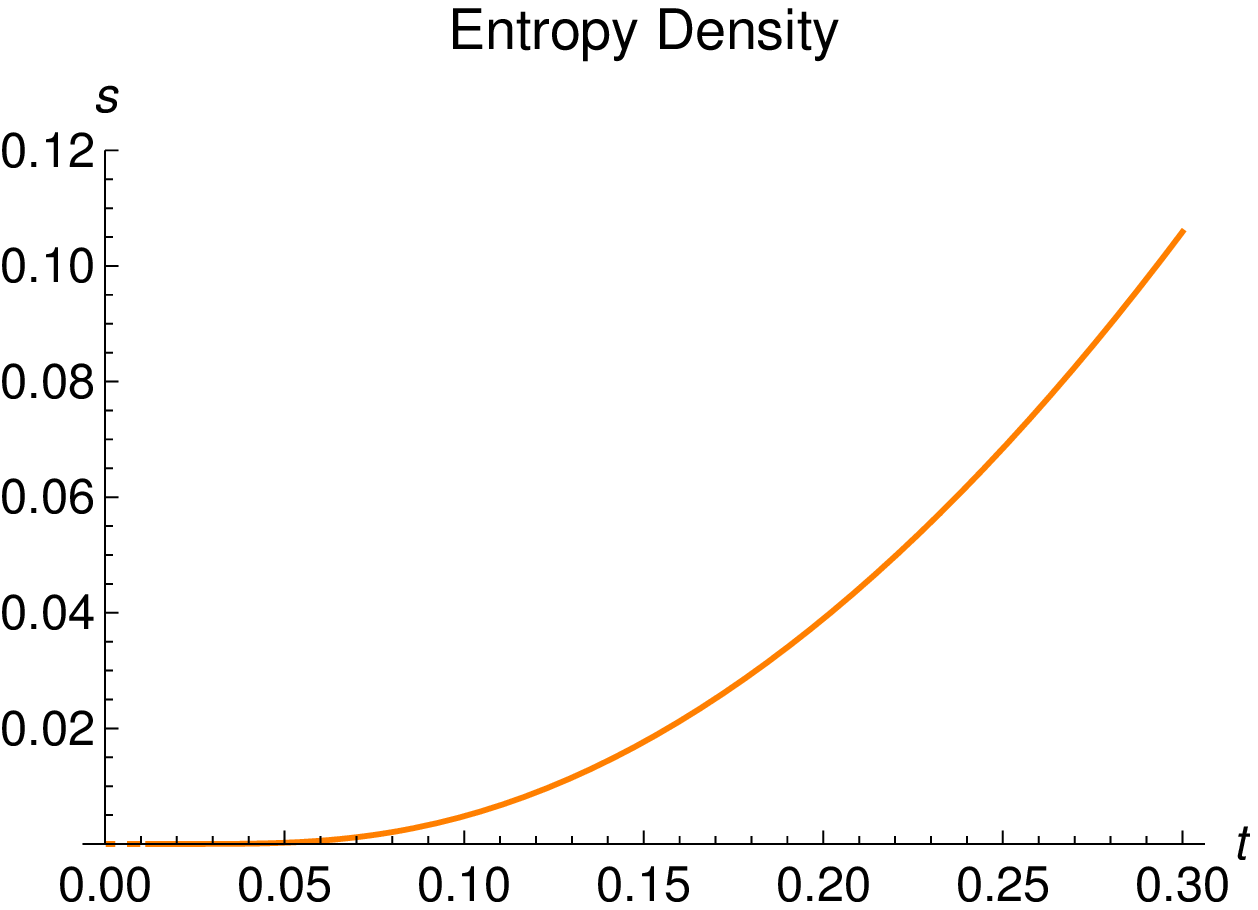}}
\vspace{4mm}
\hbox{
\includegraphics[width=7.5cm]{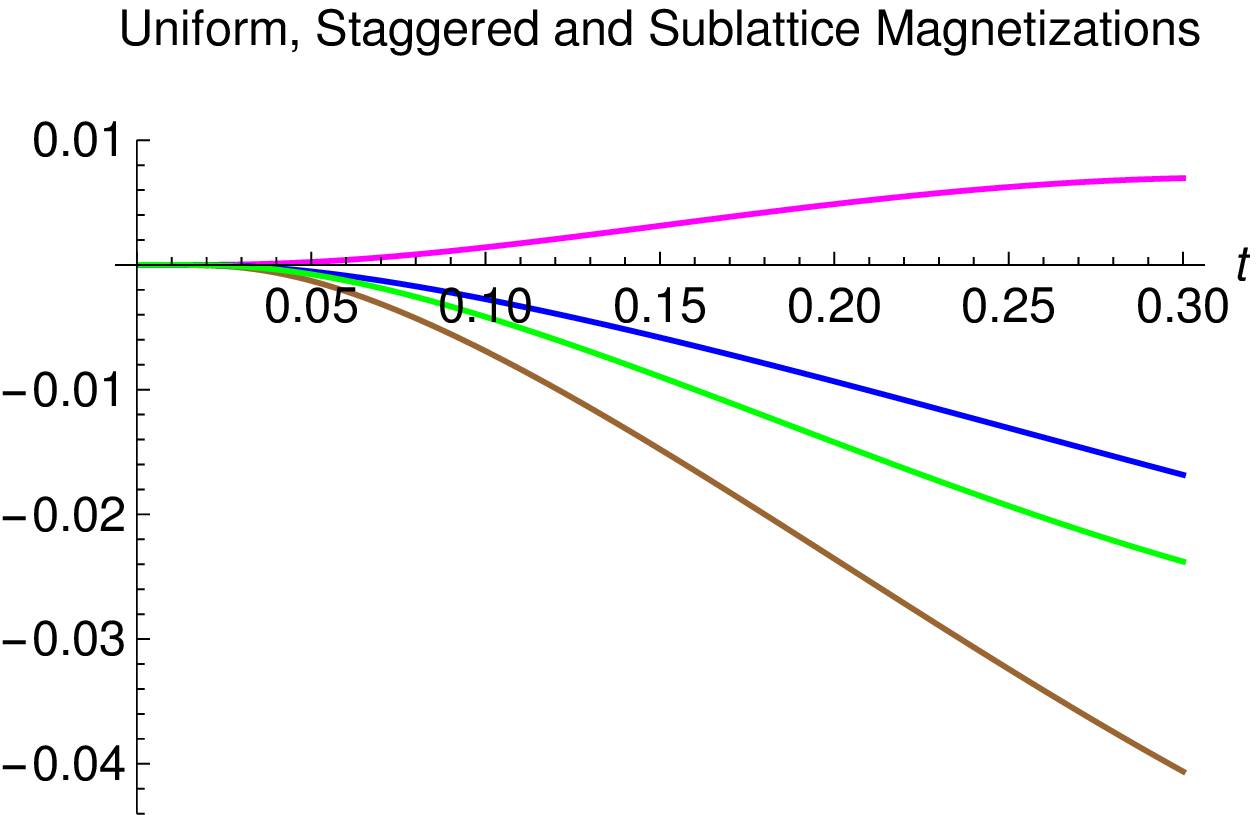}
\hspace{2mm}
\includegraphics[width=6.8cm]{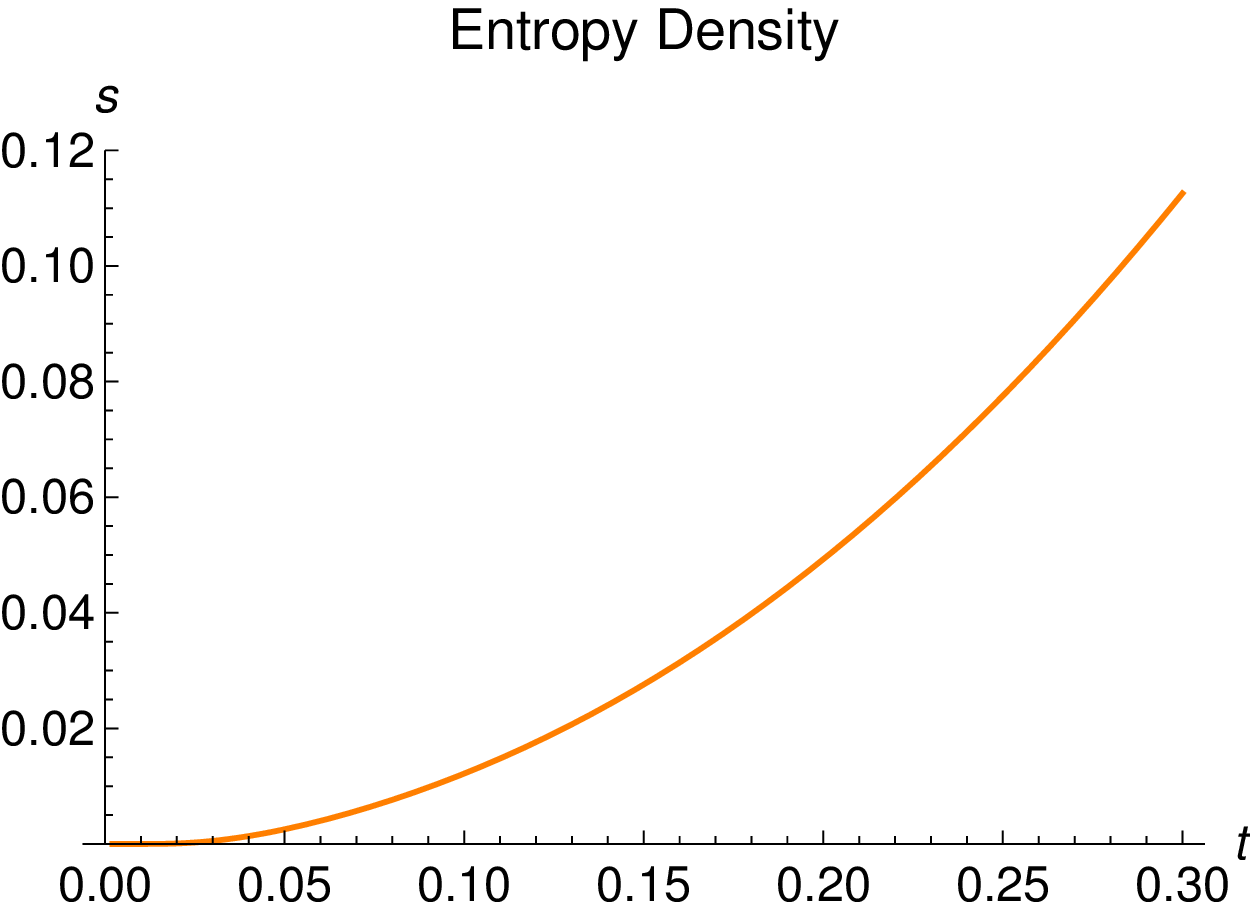}}
\end{center}
\caption{[Color online] Spin-$\frac{1}{2}$ square-lattice antiferromagnet in mutually parallel magnetic and staggered fields. LHS:
Temperature dependence of uniform (magenta), staggered (brown), as well as sublattice magnetizations $M_A$ (blue) and $|M_B|$ (green). RHS:
Temperature dependence of entropy density. Upper panel: $\{m,m_H\}=\{0.3,0.05\}$. Lower panel $\{m,m_H\}=\{0.3,0.2\}$.}
\label{figure5}
\end{figure}

First of all, if no magnetic field is present, the uniform magnetization is zero and the sublattice magnetizations $M_A$ and $|M_B|$
are identical for arbitrary temperatures and arbitrary staggered field strength. However, in presence of a magnetic field aligned with the
order parameter, interesting effects take place, as we illustrate in Fig.~\ref{figure5}. The upper panel refers to the first point
$\{m,m_H\}=\{0.3,0.05\}$ that corresponds to weak magnetic field. Remarkably, a net uniform magnetization is induced that even gets larger
at higher temperatures. This is quite counterintuitive: one would rather expect thermal fluctuations to prevent emergence of a uniform
magnetization, in particular, at more elevated temperatures. Still, as temperature continues to rise, the uniform magnetization increases
less rapidly and then eventually starts to drop (not shown in the figure).

On the other hand, the staggered magnetization diminishes when temperature rises, according to intuition. In Fig.~\ref{figure5} we
furthermore depict the sublattice magnetizations $M_A$ and $|M_B|$ that also drop when temperature increases. The essential point is that
$M_A$ and $|M_B|$ do not drop alike, such that a net uniform magnetization is created in presence of a magnetic field.

Trying to correlate these findings with the behavior of the entropy density, we note that information on microscopic order is provided by
the staggered magnetization, because it measures the total of aligned spins along the order parameter axis. Following this simple picture,
the decrease of the staggered magnetization with temperature is related to the increase of entropy with temperature:
\begin{equation}
- \frac{\mbox{d} M_s}{\mbox{d} T} \propto \frac{\mbox{d} s}{\mbox{d} T} \, .
\end{equation}
According to Fig.~\ref{figure5} this is indeed the case. For the specific point $\{m,m_H\}=\{0.3,0.05\}$, the
entropy density first remains constant, up to about $t \approx 0.07$, but then starts to increase. Likewise, staggered and sublattice
magnetization first remain constant, up to about $t \approx 0.07$, but then start to decrease. Although the counterintuitive emergence of
uniform magnetization hints at creation of spin order, the entropy increases because the destruction of spin antialignment -- reflected by
the staggered magnetization -- is the dominant effect.

Let us consider the second point $\{m,m_H\}=\{0.3,0.2\}$ that corresponds to stronger magnetic field, but still fulfills the stability
criterion, Eq.~(\ref{stabilityCondition}). According to the lower panel of Fig.~\ref{figure5}, the splitting between the curves for $M_A$
and $|M_B|$, induced by the magnetic field, is now more pronounced. Accordingly, this larger asymmetry creates a larger uniform
magnetization, as compared to the previous point $\{m,m_H\}=\{0.3,0.05\}$. The staggered magnetization and the sublattice magnetizations
$M_A$ and $|M_B|$ again decrease with temperature while the entropy density grows. The overall pattern is thus the same: the destruction of
the arrangement of antialigned spins is predominant. It should be noted that the perturbation of spin order in the stronger magnetic field
is more pronounced: compared to the previous point, the entropy here starts to increase -- and the staggered magnetization starts to
decrease -- already around the temperature $t \approx 0.03$, whereas for the previous point we had $t \approx 0.07$.

Our findings are consistent with the experimental data presented in Ref.~\citep{AUW77} where the thermomagnetic properties of the quasi
two-dimensional antiferromagnet $K_2 Mn F_4$ have been studied. The sublattice magnetizations decrease with temperature, but not alike: the
down-sublattice magnetization drops more rapidly such that a net uniform magnetization is induced that increases with temperature. In
accordance with our findings, this counterintuitive effect is more pronounced in stronger magnetic fields. Regarding the theoretical side,
we are not aware of any references -- except for Ref.~\citep{Hof20a} -- that have reported such order-disorder phenomena in
antiferromagnetic films subjected to magnetic fields aligned with {\it nonzero} staggered fields.

\section{Antiferromagnetic Films in Mutually Orthogonal Staggered and Magnetic Fields}
\label{MutuallyOrthogonal}

\subsection{Preliminaries}
\label{prelim2}

It is instructive to discuss the topic of entropy and spin order also for antiferromagnetic films that are subjected to a magnetic field
{\bf orthogonal} to the staggered field. While this general setting has been analyzed in Refs.~\citep{Hof17,Hof20b} within effective field
theory, entropy density as well as order-disorder phenomena in the staggered and uniform magnetizations have not been addressed in that
reference -- this is the objective of the present section.

The specific configuration of external fields now is
\begin{equation}
{\vec H} = (0,H,0) \, , \qquad {\vec H}_s = (H_s,0,0) \, , \qquad H, H_s > 0 \, ,
\end{equation}
i.e., magnetic and staggered fields are mutually orthogonal. The staggered field points again into the direction of the order parameter.

An essential difference with respect to the configuration of mutually parallel fields is that in the case of mutually orthogonal fields,
the two magnons satisfy the dispersion laws
\begin{eqnarray}
\label{disprelAFH}
\omega_{I} & = & \sqrt{{\vec k}^2 + \frac{M_s H_s}{\rho_s} + H^2} \, , \nonumber \\
\omega_{I\!I} & = & \sqrt{{\vec k}^2 + \frac{M_s H_s}{\rho_s}} \, ,
\end{eqnarray}
i.e., one of the magnons does not take notice of the magnetic field. As a consequence the low-energy physics is different. Readers
interested in the derivation of the corresponding free energy density are referred to Ref.~\citep{Hof17}. Here we merely quote the two-loop
result,
\begin{eqnarray}
\label{freeEDtwoLoopOrthogonal}
z & = & z_0 - \mbox{$ \frac{1}{2}$} \Big\{ g^{I}_0 + g^{I\!I}_0 \Big\} \nonumber \\
& & + \frac{M_s H_s}{16 \pi \rho^2_s} \, \Bigg\{ \sqrt{\frac{M_s H_s}{\rho_s} + H^2} - \sqrt{\frac{M_s H_s}{\rho_s}} \Bigg\} g^{I}_1
+ \frac{H^2}{4 \pi \rho_s} \, \sqrt{\frac{M_s H_s}{\rho_s} + H^2} \, g^{I}_1  \nonumber \\
& & - \frac{M_s H_s}{16 \pi \rho^2_s} \, \Bigg\{ \sqrt{\frac{M_s H_s}{\rho_s} + H^2} - \sqrt{\frac{M_s H_s}{\rho_s}} \Bigg\} g^{I\!I}_1 \\
& & - \frac{M_s H_s}{8 \rho^2_s} \, \Big\{ {(g^{I}_1)}^2 - 2 g^{I}_1 g^{I\!I}_1 + {(g^{I\!I}_1)}^2 \Big\}
- \frac{H^2}{2 \rho_s}{(g^{I}_1)}^2 + \frac{2}{\rho_s} \, s(\sigma,\sigma_H) \, T^4 \, , \nonumber
\end{eqnarray}
where the vacuum energy density $z_0$ is
\begin{eqnarray}
\label{vacuumEDtwoLoopOrthogonal}
z_0 & = & - M_s H_s - \mbox{$ \frac{1}{2}$} \rho_s H^2
- (k_2 + k_3) \frac{M^2_s H^2_s}{\rho^2_s} - k_1 \frac{M_s H_s}{\rho_s} H^2 -(e_1 + e_2) H^4 \nonumber \\
& & - \frac{1}{12 \pi} \Bigg\{ {\Big( \frac{M_s H_s}{\rho_s} + H^2 \Big) }^{3/2} + {\Big( \frac{M_s H_s}{\rho_s} \Big)}^{3/2} \Bigg\}
- \frac{M_s^2 H_s^2}{64 \pi^2 \rho^3_s} \nonumber \\
& & - \frac{5 M_s H_s H^2}{128 \pi^2 \rho^2_s} - \frac{H^4}{32 \pi^2 \rho_s}
+ \frac{M_s^{3/2} H_s^{3/2}}{64 \pi^2 \rho^{5/2}_s} \, \sqrt{\frac{M_s H_s}{\rho_s} + H^2} \, ,
\end{eqnarray}
and the kinematical Bose functions $g^{I,{I\!I}}_i$ and the sunset function $s(\sigma,\sigma_H)$ are provided in appendix \ref{appendixA}. The
vacuum energy density -- in addition to $k_2$ and $k_3$ that already showed up in the previous case of mutually parallel fields -- contains
the next-to-leading order effective constants $e_1, e_2, k_1$.\footnote{All relevant NLO effective constants -- $k_1,k_2,k_3,e_1,e_2$ -- are
defined in section 2 of Ref.~\citep{Hof20a}.}

\subsection{Entropy Density}
\label{entropyMutuallyOrthogonal}

Let us first look at the entropy density that one readily derives from the two-loop free energy density,
Eq.~(\ref{freeEDtwoLoopOrthogonal}), as
\begin{eqnarray}
\label{entropyTwoLoopOrthogonal}
s(t,m,m_H) & = & \frac{1}{2} \, \Bigg( \frac{\mbox{d} g^{I}_0}{\mbox{d}T} + \frac{\mbox{d} g^{I\!I}_0}{\mbox{d}T} \Bigg)
- \frac{\pi^2 \rho_s^2 m^2}{2} \, \Big( \sqrt{m^2 + m^2_H} - m \Big) \frac{\mbox{d} g^I_1}{\mbox{d}T} \nonumber \\
& & - \pi^2 \rho_s^2 m^2_H \, \sqrt{m^2 + m^2_H}  \, \frac{\mbox{d} g^I_1}{\mbox{d}T}
+ \frac{\pi^2 \rho_s^2 m^2}{2} \, \Big( \sqrt{m^2 + m^2_H} - m \Big) \frac{\mbox{d} g^{I\!I}_1}{\mbox{d}T} \nonumber \\
& & + 2 \pi^3 \rho_s^2 m^2 \Bigg( g^I_1 \, \frac{\mbox{d} g^I_1}{\mbox{d}T}
+ g^{I\!I}_1  \, \frac{\mbox{d} g^{I\!I}_1}{\mbox{d}T}
- g^{I\!I}_1  \, \frac{\mbox{d} g^I_1}{\mbox{d}T}
- g^I_1  \, \frac{\mbox{d} g^{I\!I}_1}{\mbox{d}T} \Bigg) \nonumber \\
& & + 4 \pi^2 \rho_s m^2_H g^I_1  \, \frac{\mbox{d} g^I_1}{\mbox{d}T}
- \frac{2}{\rho_s} \frac{\mbox{d} s(\sigma,\sigma_H)}{\mbox{d}T} \, T^4
- \frac{8}{\rho_s} \, s(\sigma,\sigma_H) \, T^3 \, .
\end{eqnarray}
It should be stressed that next-to-leading order effective constants only show up in the (zero-temperature) vacuum energy density $z_0$.
The finite-temperature behavior of the system, up to two-loop order, is again fully described in terms of the leading-order effective
constants $\rho_s$ and $M_s$. In the representation for the entropy density, $M_s$ is hidden in the low-energy parameter $m$.

\begin{figure}
\begin{center}
\hbox{
\includegraphics[width=7.5cm]{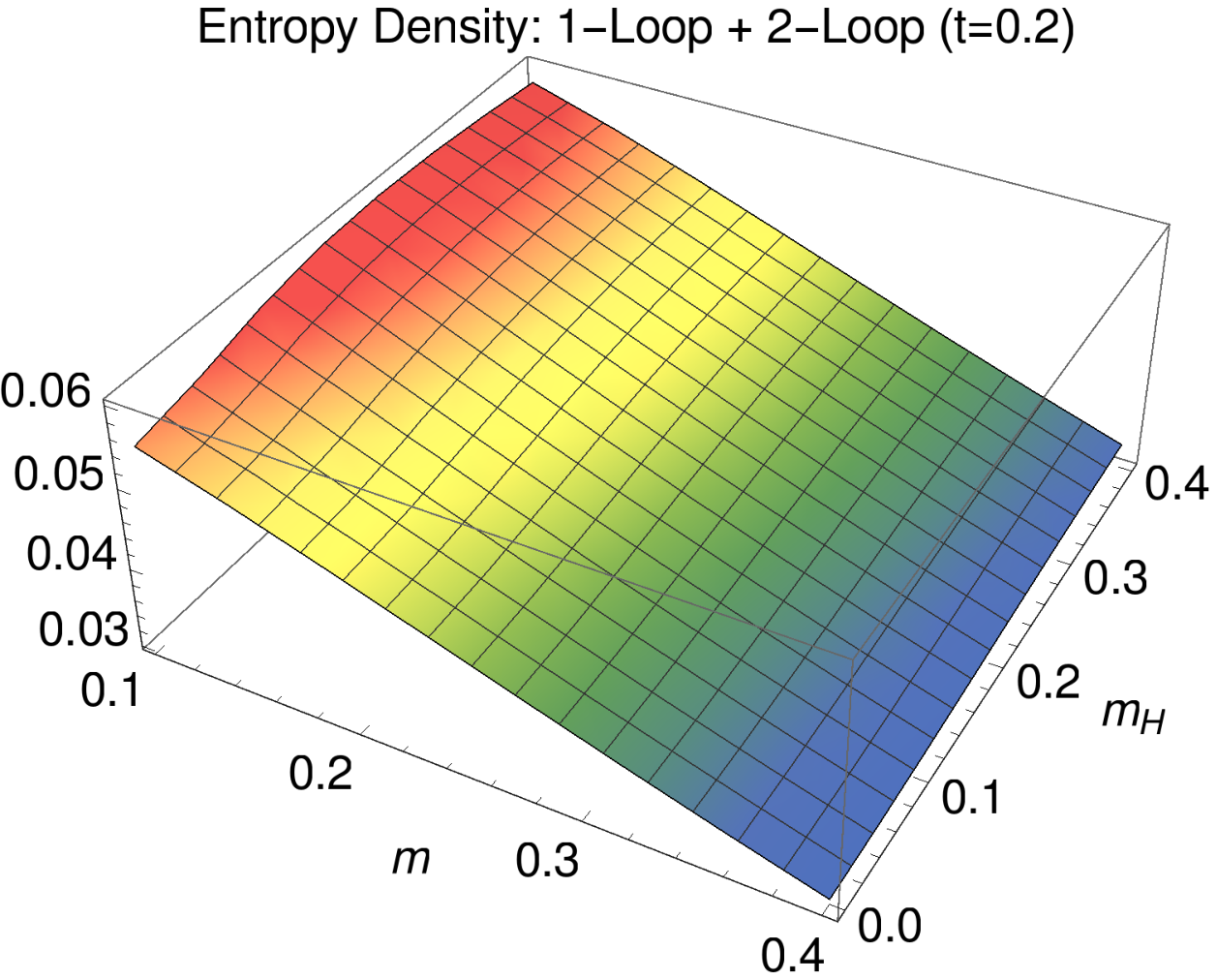} 
\hspace{2mm}
\includegraphics[width=7.5cm]{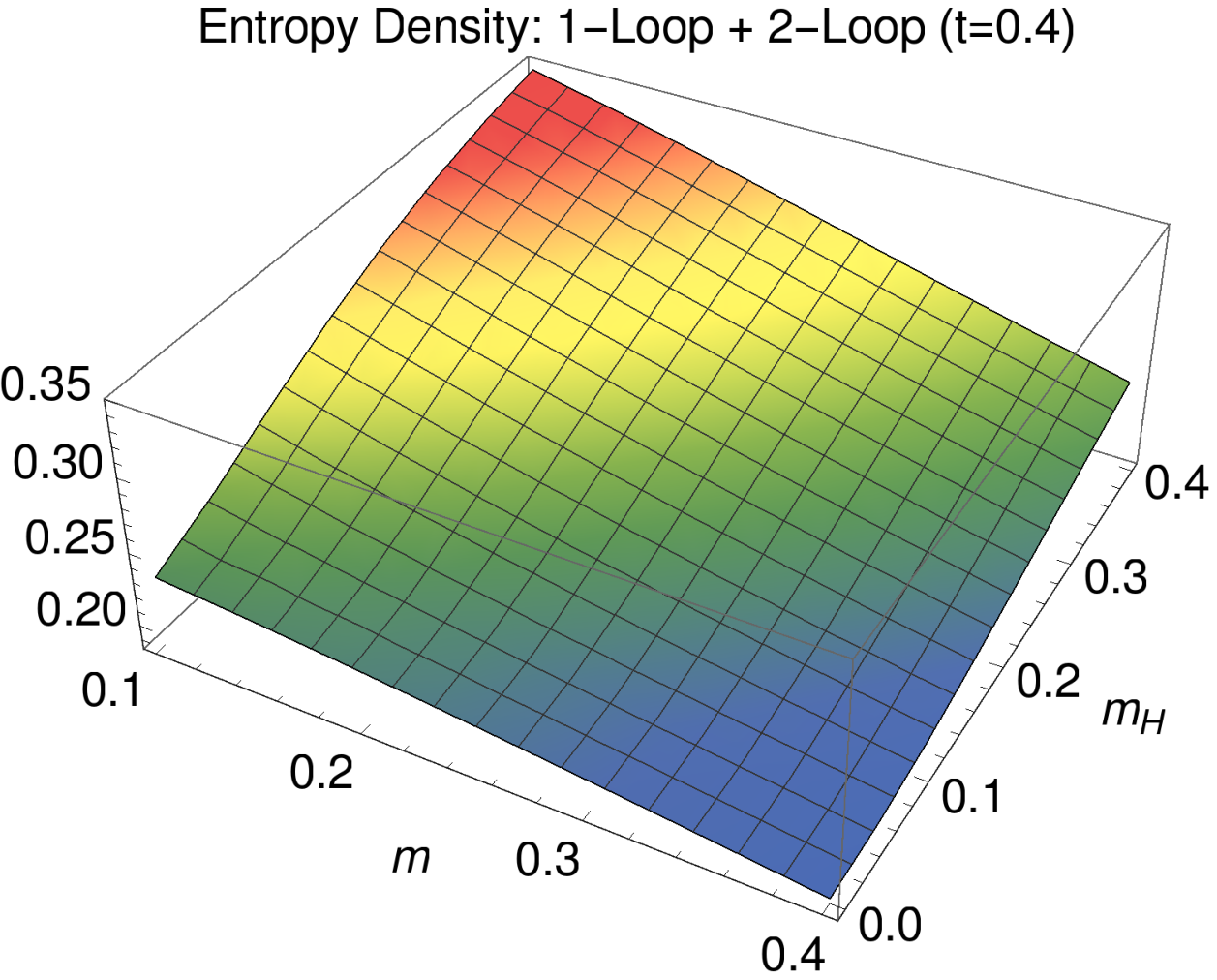}}
\end{center}
\caption{[Color online] Entropy density $s$ for the spin-$\frac{1}{2}$ square-lattice antiferromagnet in mutually orthogonal magnetic
($m_H$) and staggered ($m$) fields at the temperatures $t=0.2$ and $t=0.4$.}
\label{figure6}
\end{figure}

In analogy to the plots in the previous section, in Fig.~\ref{figure6}, we consider the entropy density $s(t,m,m_H)$ of the
spin-$\frac{1}{2}$ square-lattice antiferromagnet as a function of magnetic and staggered field strength for the two temperatures $t=0.2$
and $t=0.4$. Since here no stability criterion applies, we also consider the parameter region that was excluded in the case of mutually
parallel fields. Overall, the entropy drops when the staggered field grows, and the entropy increases when the magnetic field grows. This
behavior is qualitatively the same as for mutually parallel fields and naively one concludes again that the staggered field establishes
spin order by enforcing antialignment of the spins, while the magnetic field perturbs the antialigned array of spins. 

\subsection{Order-Disorder Effects}
\label{SubMagMutuallyOrthogonal}

We first comment on a qualitative difference between antiferromagnetic films in mutually orthogonal and mutually parallel fields. When the
fields are parallel, the relative strengths of magnetic and staggered field cannot take arbitrary values -- rather a stability criterion
must be obeyed. Moreover, uniform, staggered and sublattice magnetizations are all measured with respect to the same axis that is defined
by the staggered magnetization vector at zero temperature, i.e., the order parameter.

On the other hand, if magnetic and staggered fields are orthogonal, the magnetic field can take larger values than the staggered field --
no stability criterion must be met. Also, while staggered and sublattice magnetizations are measured with respect to the order parameter
axis, the uniform magnetization that is induced by the external magnetic field, points orthogonal to the order parameter. In such a
configuration of external fields, no uniform magnetization in the direction of the order parameter can be created by the magnetic field
whose effect rather is to tilt the spins in its proper direction. Magnetic and staggered fields hence affect sublattice $A$ and sublattice
$B$ in the same way: it makes no sense in this case to produce individual plots for the sublattice magnetizations whose magnitude simply is
half of the staggered magnetization.

To begin with, we analyze the situation at zero temperature, where staggered and uniform magnetizations are given by (see
Ref.~\citep{Hof17})
\begin{eqnarray}
\label{OPorthogonalT0}
\frac{M_s(0,m,m_H)}{M_s} & = & 1 + \frac{m}{4} + \frac{\sqrt{m^2+m_H^2}}{4} + \frac{m^2}{8} + \frac{5 m_H^2}{32}
- \frac{m^3}{8 \sqrt{m^2+m_H^2}} \nonumber \\
& & - \frac{3 m \, m_H^2}{32 \sqrt{m^2+m_H^2}} + 8 \pi^2 \rho_s (k_2 + k_3) \, m^2 + 4 \pi^2 \rho_s k_1 \, m_H^2 \, , \nonumber \\
& & \hspace{-2.4cm} m = \frac{\sqrt{M_s H_s}}{2 \pi \rho^{3/2}_s} \, , \qquad m_H = \frac{H}{2 \pi \rho_s} \, , \qquad 
M_s = M_s(0,0,0) \, ,
\end{eqnarray}
and
\begin{eqnarray}
\label{MagorthogonalT0}
\frac{M(0,m,m_H)}{\rho^2_s} & = & 2 \pi \, m_H + \pi \, m_H \sqrt{m^2+m_H^2} + \pi m_H^3 + \frac{5 \pi}{8} \, m^2 \, m_H \\
& &- \frac{\pi}{8} \, \frac{m^3 \, m_H}{\sqrt{m^2+m_H^2}} + 32 \pi^3 \rho_s (e_1 + e_2) \, m_H^3 + 16 \pi^3 \rho_s k_1 \, m^2 m_H \, ,
\nonumber
\end{eqnarray}
respectively. The staggered magnetization, in addition to the combination $k_2 + k_3$ of NLO effective constants that showed up previously
in Eq.~(\ref{OPT0}), here also depends on $k_1$. Moreover, the uniform magnetization involves the NLO effective constants $e_1$ and $e_2$
that were absent in the case of mutually parallel fields, Eq.~(\ref{MagT0}). Unlike for the combination $k_2 + k_3$, the numerical values
of $k_1, e_1, e_2$ are not available from Monte Carlo simulations. Estimates based on scaling arguments given in Ref.~\citep{GJMM16,Hof16}
however show that their absolute values are small -- of the order of $0.001 \, \rho^{-1}_s$ -- while their respective signs remain open.
Since we are dealing with only minor corrections, in the following figure referring to $T$=0, we neglect these contributions. As we will
see, at $T \neq 0$, all these NLO effective constants are absent, such that the finite-temperature results for $M_s$ and $M$ -- much like
the entropy density, Eq.~(\ref{entropyTwoLoopOrthogonal}) -- are parameter free. 

\begin{figure}
\begin{center}
\hbox{
\includegraphics[width=7.5cm]{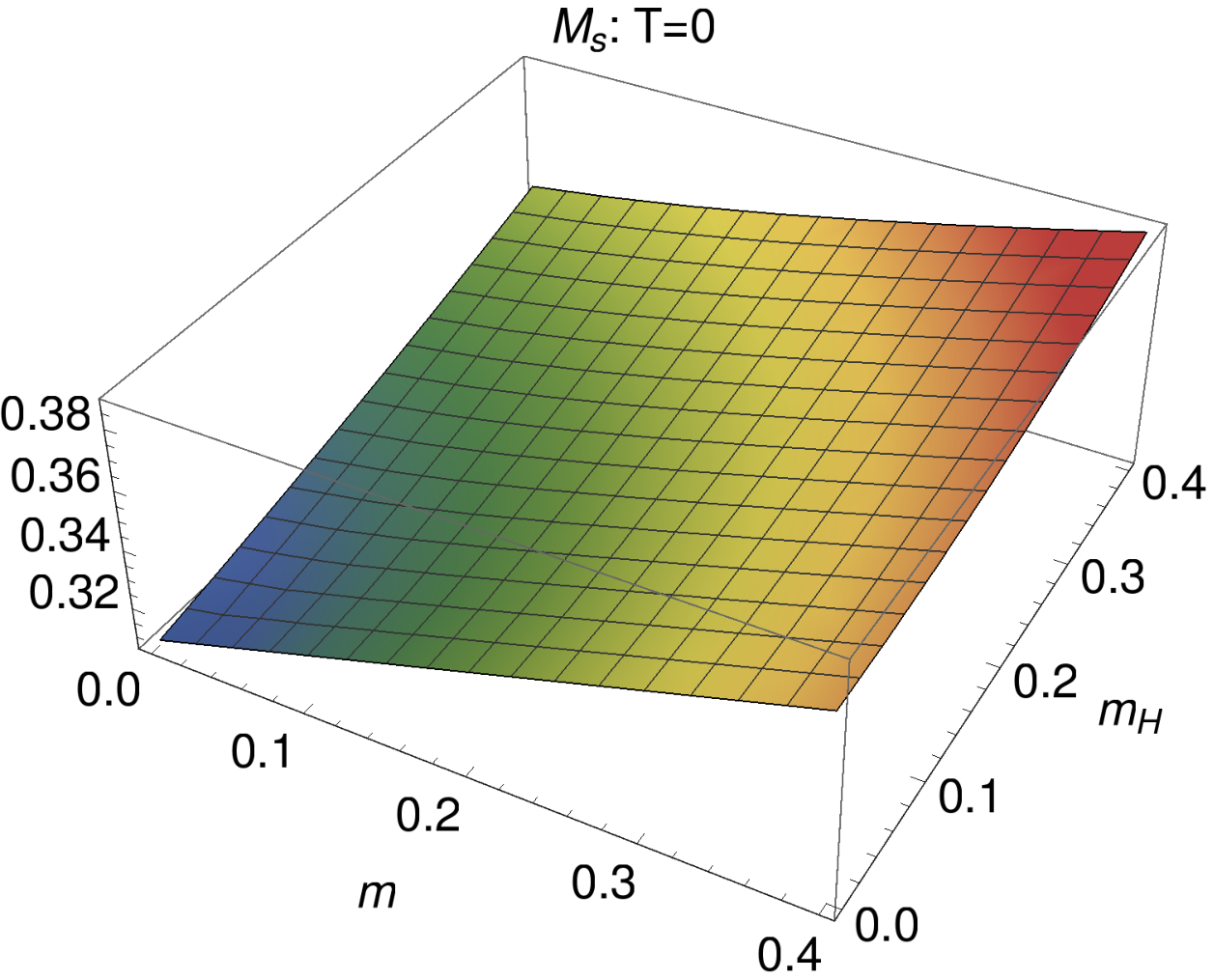}
\hspace{2mm}
\includegraphics[width=7.5cm]{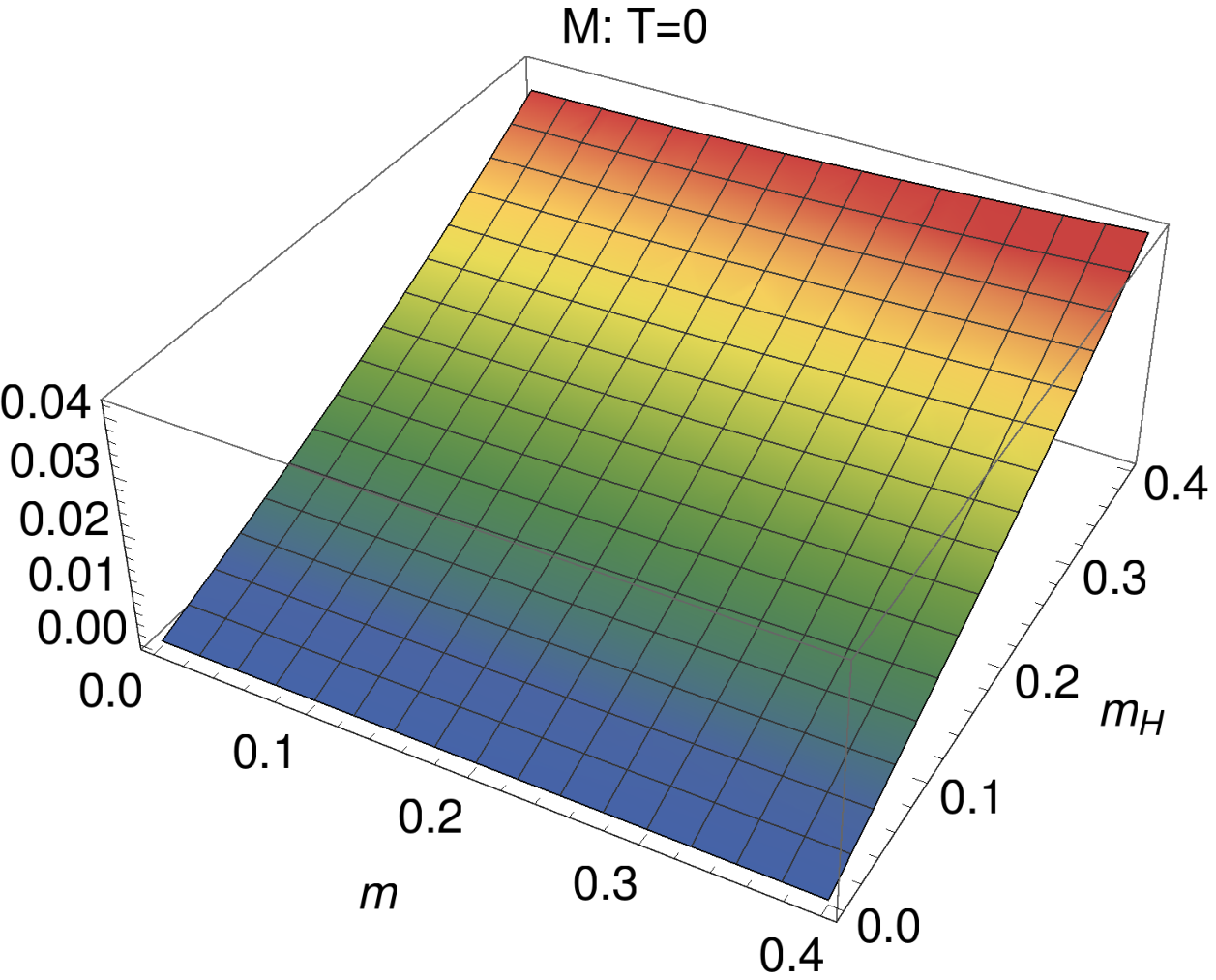}}
\end{center}
\caption{[Color online] Staggered magnetization $M_s$ and uniform magnetization $M$ for the spin-$\frac{1}{2}$ square-lattice
antiferromagnet in mutually orthogonal magnetic ($m_H$) and staggered ($m$) fields at $T$=0.}
\label{figure7}
\end{figure}

In Fig.~\ref{figure7}, we depict the dependence of the staggered and uniform magnetizations on $m$ and $m_H$ at $T$=0. The staggered
magnetization $M_s(0,m,m_H)$ grows when the magnetic or the staggered field become stronger -- much like in the previous case of mutually
parallel external fields. The enhancement of the order parameter can be explained again via suppression of quantum fluctuations caused by
the external fields. Although here the magnetic field points orthogonal to the staggered magnetization vector, it also suppresses quantum
fluctuations. The overall suppression of quantum fluctuations is of the same magnitude as in the case of mutually parallel fields:
following Fig.~\ref{figure2}, the increase in the sublattice magnetizations (strong fields compared to weak fields) is about
$0.035 \, a^{-2}$ which adds up to $0.07 \, a^{-2}$ for the staggered magnetization -- this is about the same change one observes in the
staggered magnetization in Fig.~\ref{figure7}. Finally, the uniform magnetization $M(0,m,m_H)$ -- as expected -- is zero when $H=0$, but
continuously grows when the magnetic field strength increases. The dependence on the staggered field is tiny, but it also slightly
suppresses quantum fluctuations.

\begin{figure}
\begin{center}
\hbox{
\includegraphics[width=7.3cm]{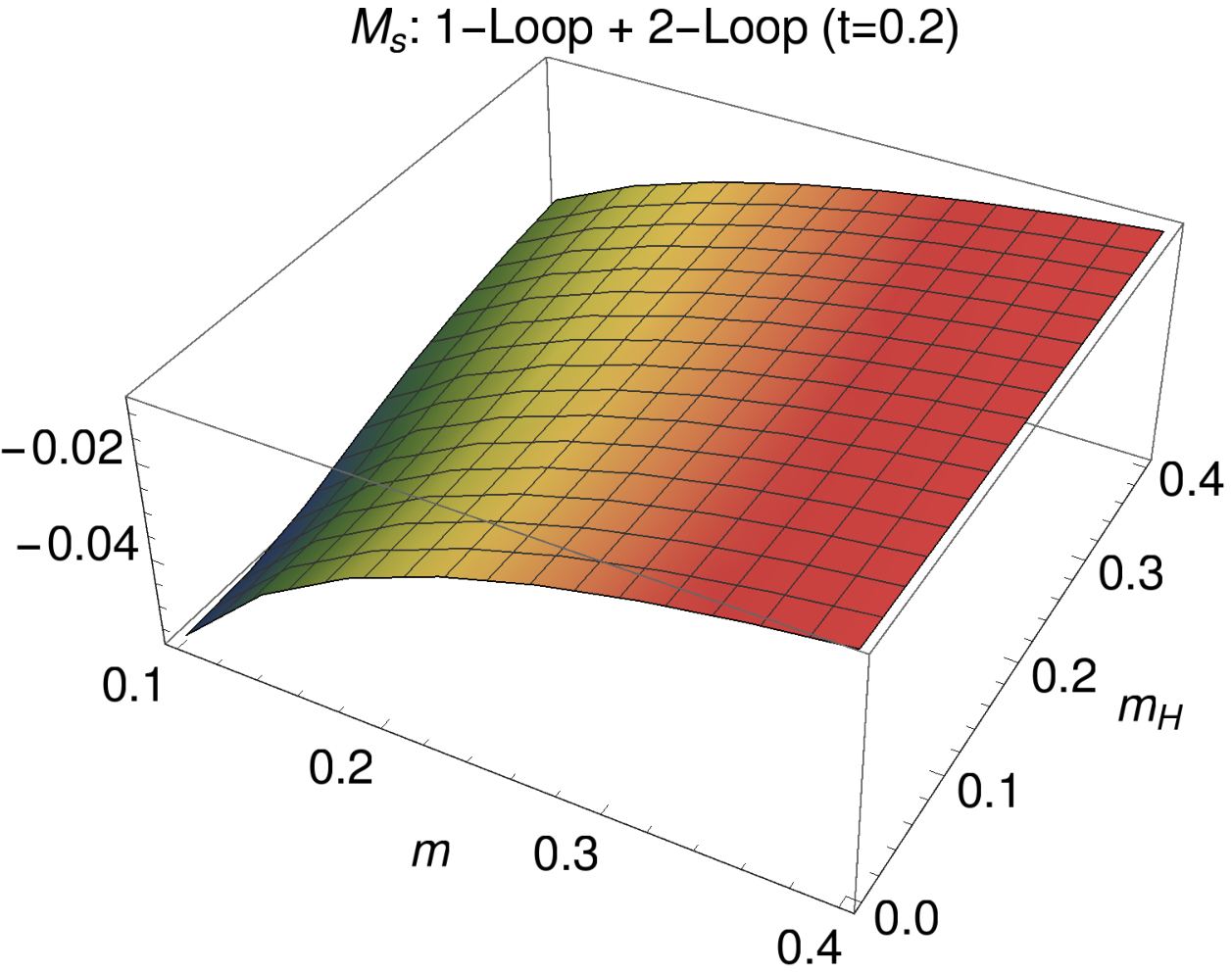}
\hspace{2mm}
\includegraphics[width=7.3cm]{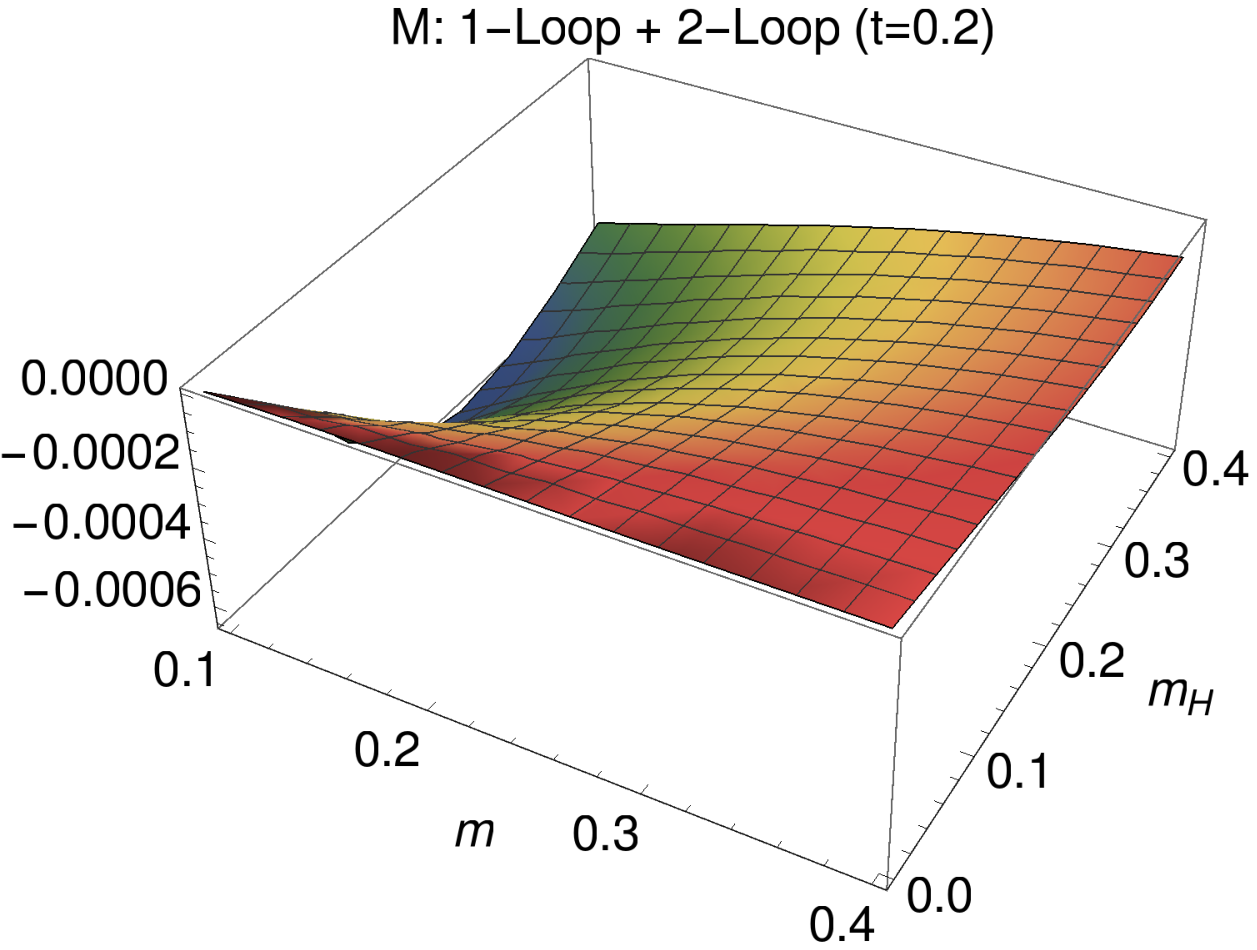}}
\vspace{2mm}
\hbox{
\includegraphics[width=7.3cm]{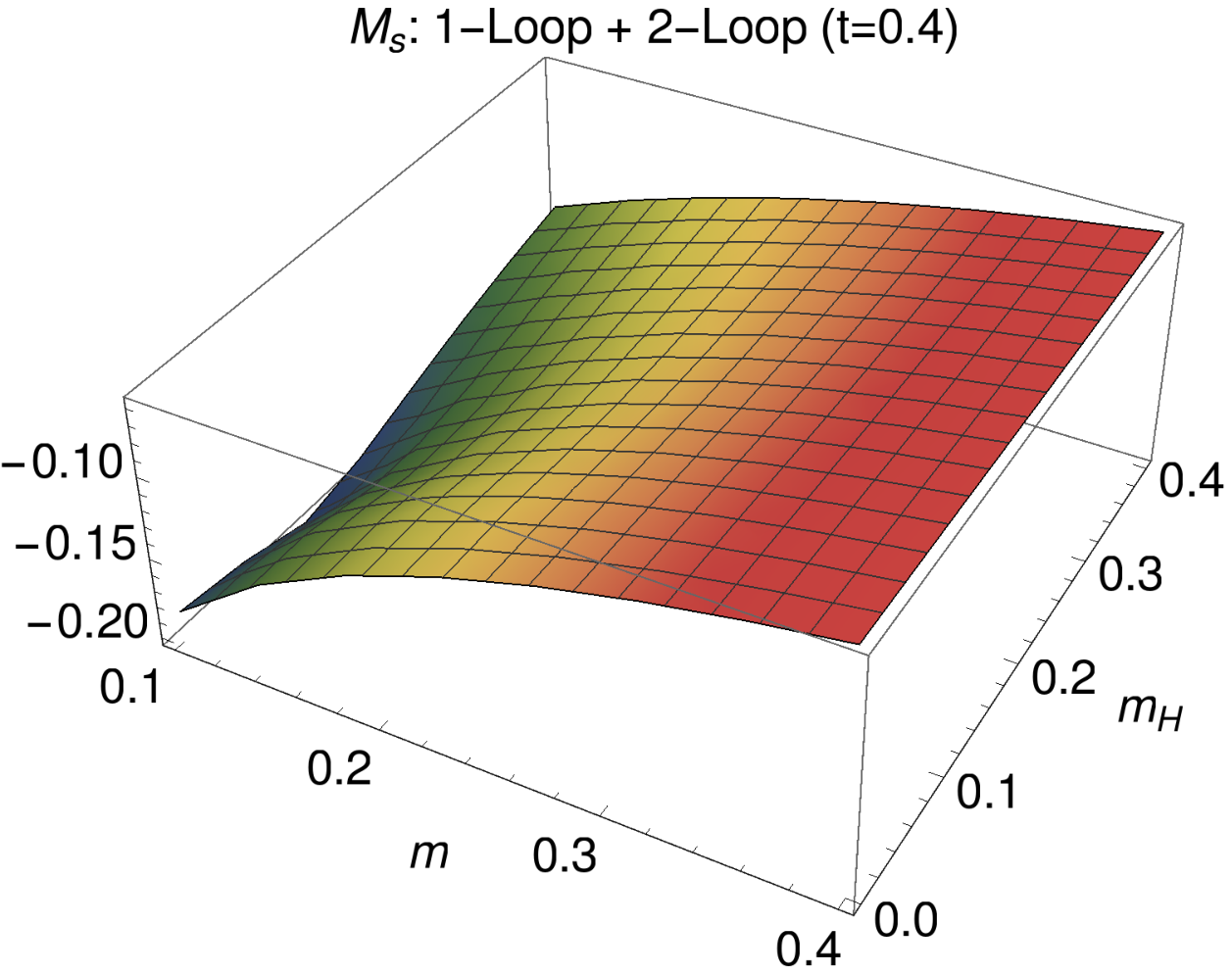}
\hspace{2mm}
\includegraphics[width=7.3cm]{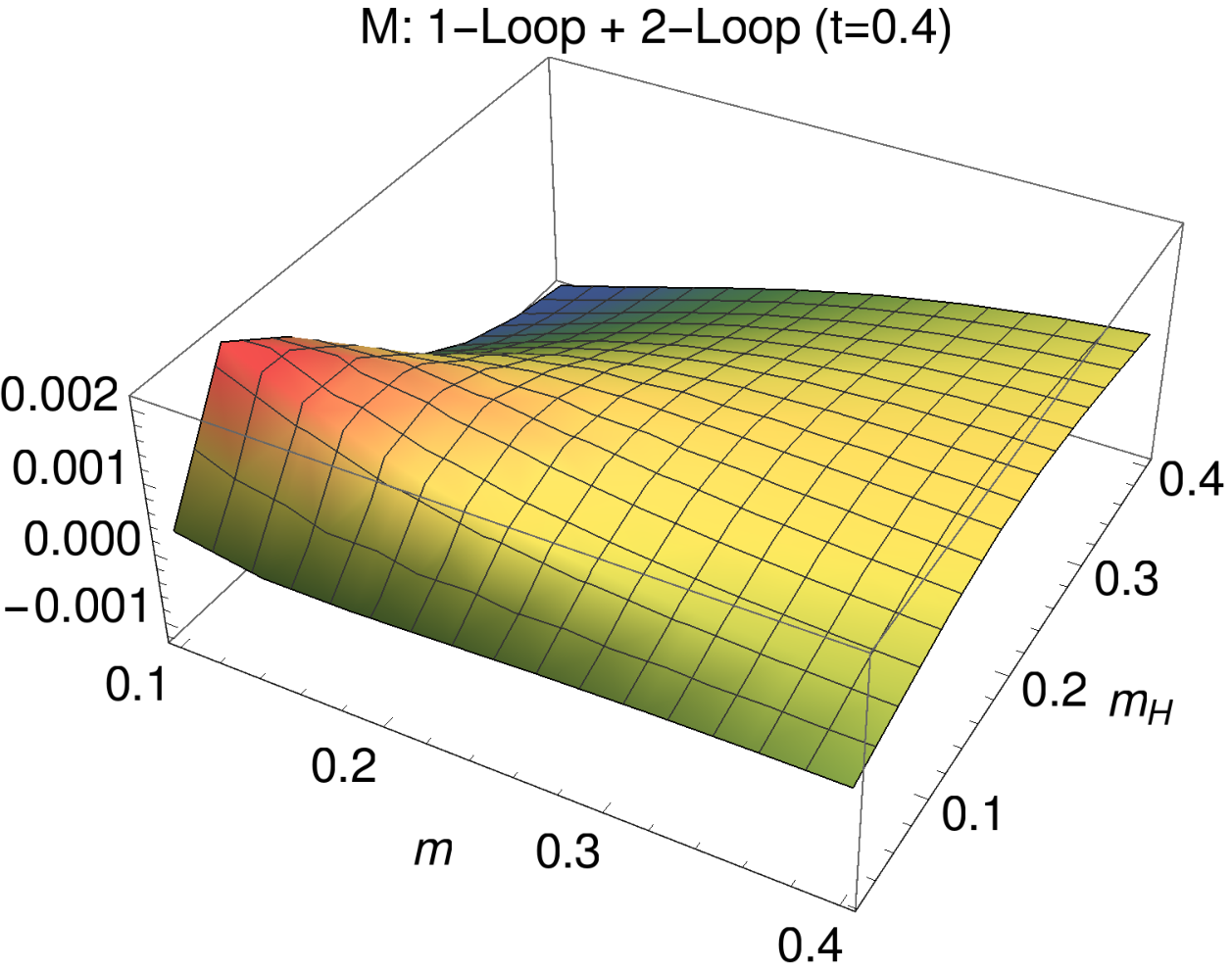}}
\end{center}
\caption{[Color online] Staggered and uniform magnetization for the spin-$\frac{1}{2}$ square-lattice antiferromagnet at the temperatures
$t = 0.2$ (upper panel) and $t = 0.4$ (lower panel): Dependence on mutually orthogonal magnetic ($m_H$) and staggered ($m$) fields.}
\label{figure8}
\end{figure}

Let us now turn to finite temperature where the expansions for the staggered and uniform magnetization take the form
\begin{eqnarray}
\label{OPMagOrthogonal}
M_s(t,m,m_H) & = & M_s(0,m,m_H) + {\tilde \sigma}_1 T + {\tilde \sigma}_2 T^2 + {\cal O}(T^3) \, , \nonumber \\
M(t,m,m_H) & = & M(0,m,m_H) + {\hat \sigma}_1 T + {\hat \sigma}_2 T^2 + {\cal O}(T^3) \, ,
\end{eqnarray}
with
\begin{eqnarray}
{\tilde \sigma}_1(t,m,m_H) & = & -\frac{M_s}{2 \rho_s} \, \Big( h^{I}_1 + h^{I\!I}_1 \Big) \, , \nonumber \\
{\hat \sigma}_1(t,m,m_H) & = & - 2 \pi \rho_s m_H h^{I}_1 \, .
\end{eqnarray}
We refrain from listing the rather lengthy expressions for the coefficients ${\tilde \sigma}_2$ and ${\hat \sigma}_2$ that can be obtained
trivially from the free energy density, Eq.(\ref{freeEDtwoLoopOrthogonal}), via
\begin{eqnarray}
M_s(T,H_s,H) & = & - \frac{\partial z(T,H_s,H)}{\partial H_s} \, , \nonumber \\
M(T,H_s,H) & = & - \frac{\partial z(T,H_s,H)}{\partial H} \, .
\end{eqnarray}
Plots for the staggered and the uniform magnetization as functions of $m_H$ and $m$ for the temperatures $t=0.2$ and $t=0.4$ are provided
in Fig.~\ref{figure8}. We depict again the sum of 1-loop and 2-loop contributions given in Eq.~(\ref{OPMagOrthogonal}),
\begin{eqnarray}
& M_s: &  {\tilde \sigma}_1 T + {\tilde \sigma}_2 T^2 \, , \nonumber \\
& M: &  {\hat \sigma}_1 T + {\hat \sigma}_2 T^2 \, ,
\end{eqnarray}
without superimposing the dominant $T$=0 contributions. The plots thus capture the change of staggered and uniform magnetizations when
temperature is raised from $t$=0 to $t=\{0.2, 0.4\}$. The change in the staggered magnetization, as expected, is negative: it drops on
account of the thermal fluctuations and the decrease is most drastic in weak fields where the thermal disruption of spin order is
strongest. This is consistent with the previous observation (see Fig.~\ref{figure6}) that the entropy density drops when the staggered
field becomes stronger: the antialigned spin pattern is enforced. Unlike in mutually parallel external fields (compare with
Fig.~\ref{figure3}), at more elevated temperatures ($t=0.4$) and in stronger external fields, the effect of the magnetic field is almost
nil.

The change in the uniform magnetization, when temperature is raised from $t$=0 to $t=0.2$, is negative -- as one would expect. Remarkably,
when temperature is raised from $t$=0 to the more elevated temperature $t=0.4$, things are less intuitive: the uniform magnetization may be
enhanced or damped. In particular, the observation that the enhancement takes place in weaker external fields is quite counterintuitive.
Note however that the effects induced in the uniform magnetization are minor compared to the effects in the staggered magnetization: there
is about a two order of magnitude difference between the changes in $M_s$ and $M$ in the respective plots of Fig.~\ref{figure8}. Also,
compared to the case of mutually parallel fields (see RHS of Fig.~\ref{figure3}), the effects in the uniform magnetization are much less
pronounced here: it is more difficult to induce a net magnetization when the magnetic field points orthogonal to the order parameter. 

Let us finally explore in more detail how entropy and magnetizations vary with temperature. To this end we consider the same two
representative points in parameter space $\{m,m_H\}=\{0.3,0.05\}$ and $\{m,m_H\}=\{0.3,0.2\}$, respectively, and evaluate entropy density,
uniform magnetization and staggered magnetization as functions of temperature. The quantities $M_s$ and $M$ again refer to the
magnetizations induced by finite temperature, i.e., correspond to the response of the system when temperature is raised from $t=0$ to
$t \neq 0$.

\begin{figure}
\begin{center}
\hbox{
\includegraphics[width=7.3cm]{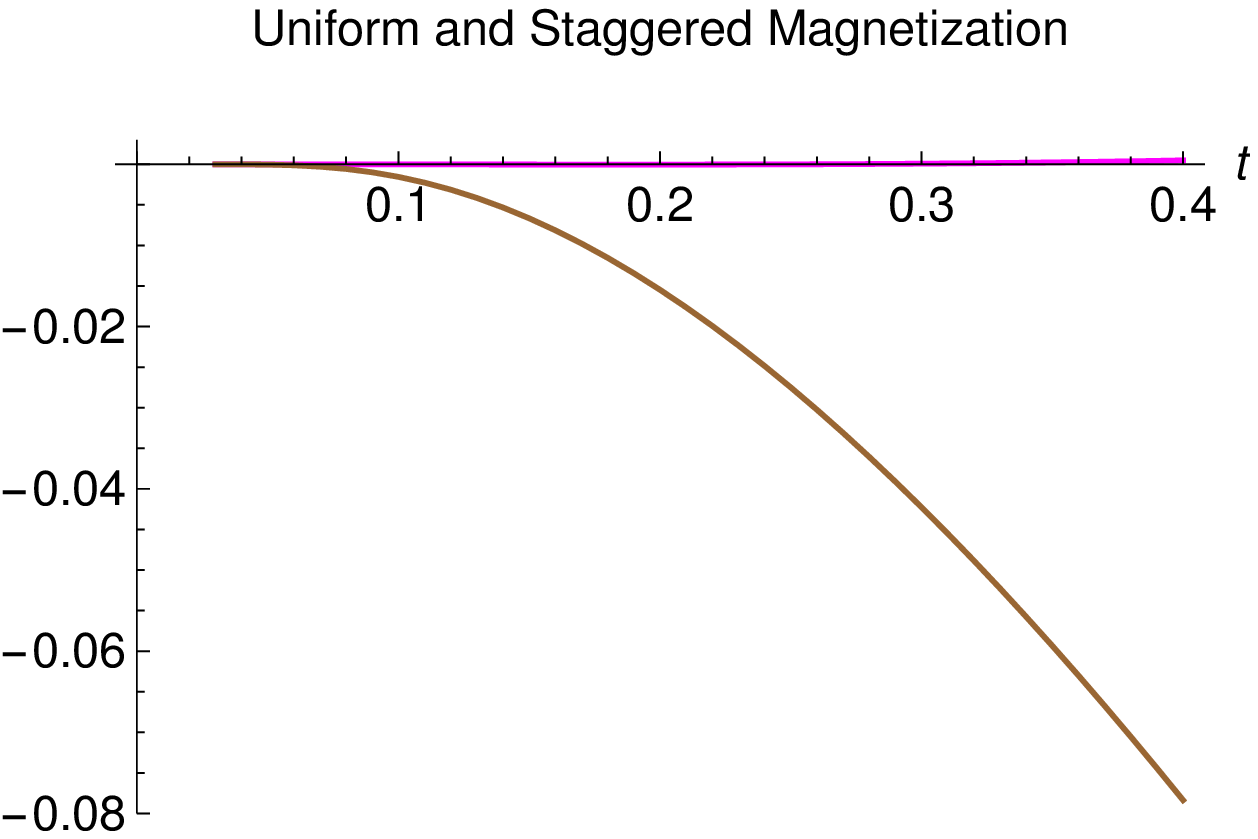}
\hspace{2mm}
\includegraphics[width=6.8cm]{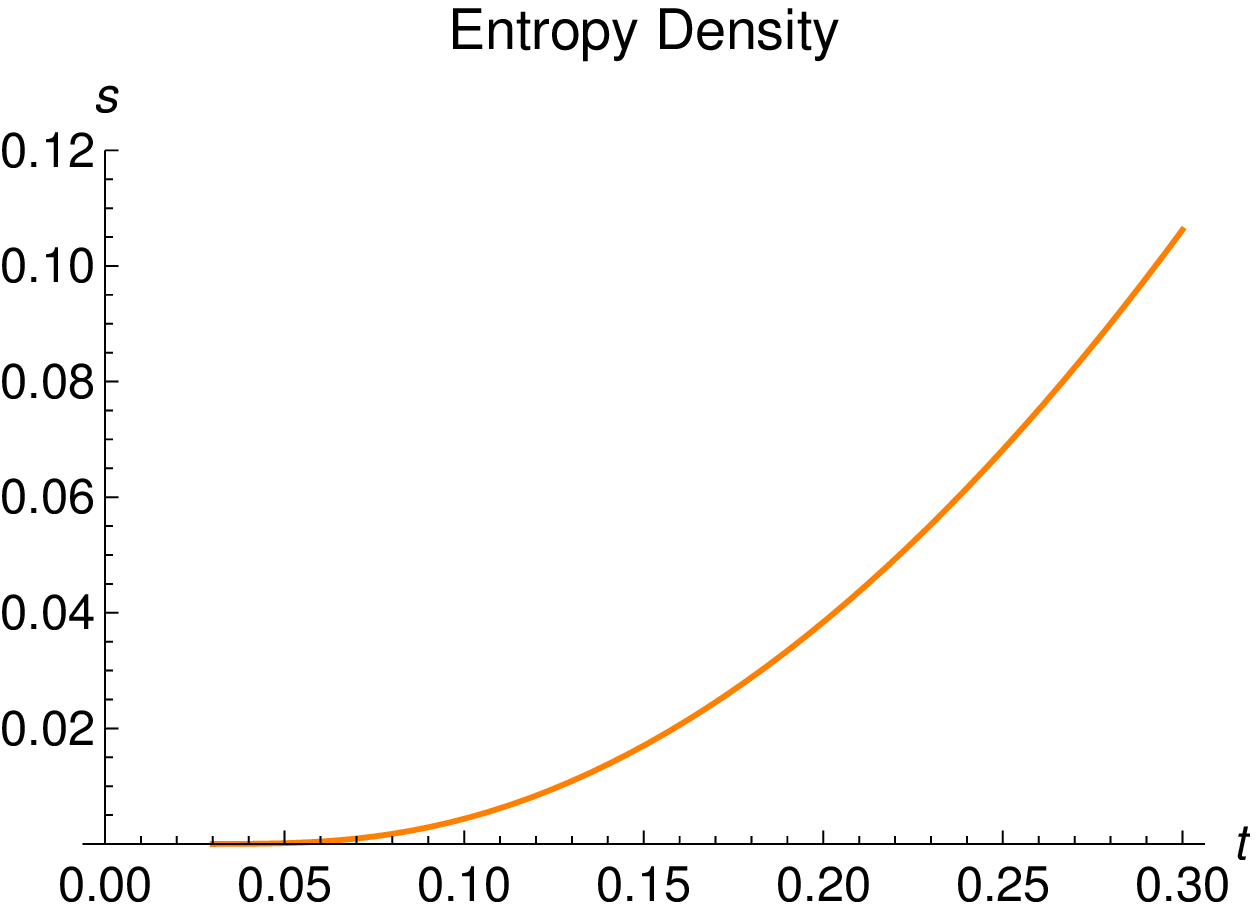}}
\vspace{4mm}
\hbox{
\includegraphics[width=7.3cm]{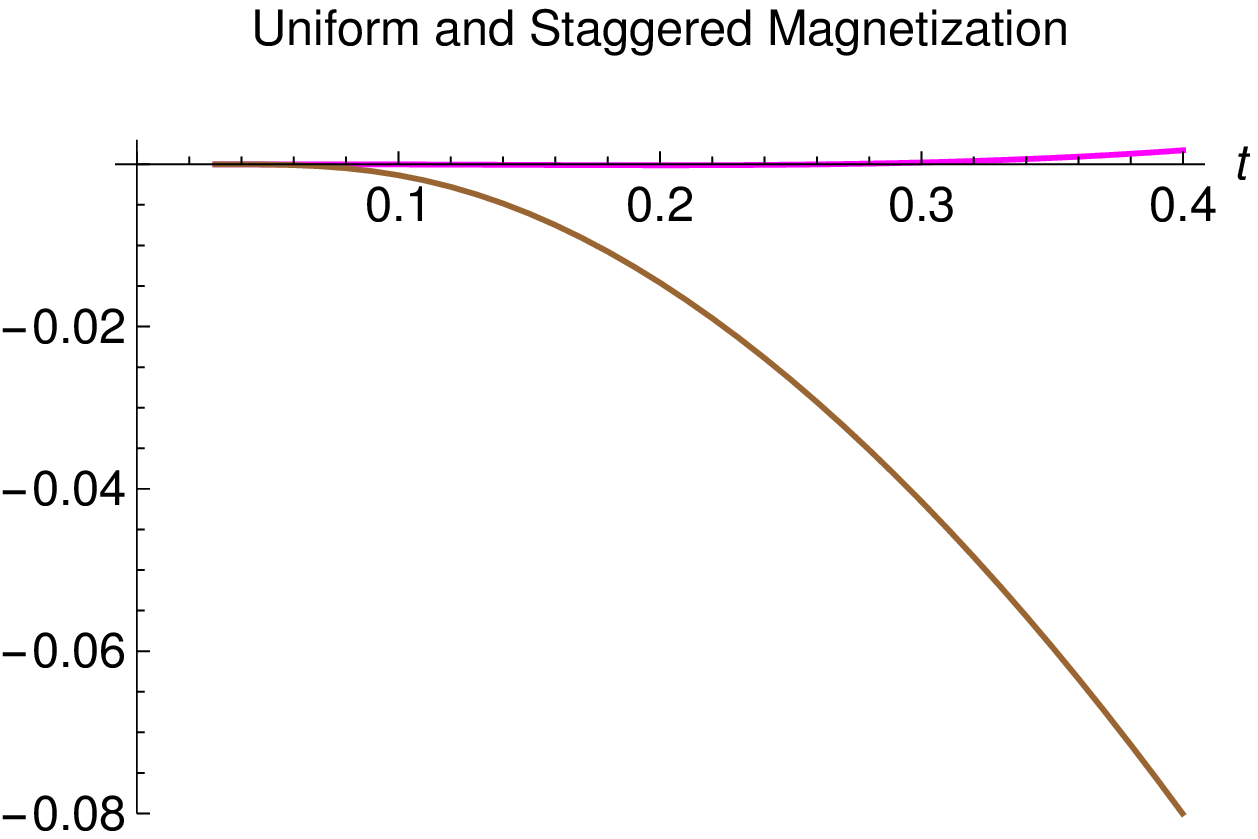}
\hspace{2mm}
\includegraphics[width=6.8cm]{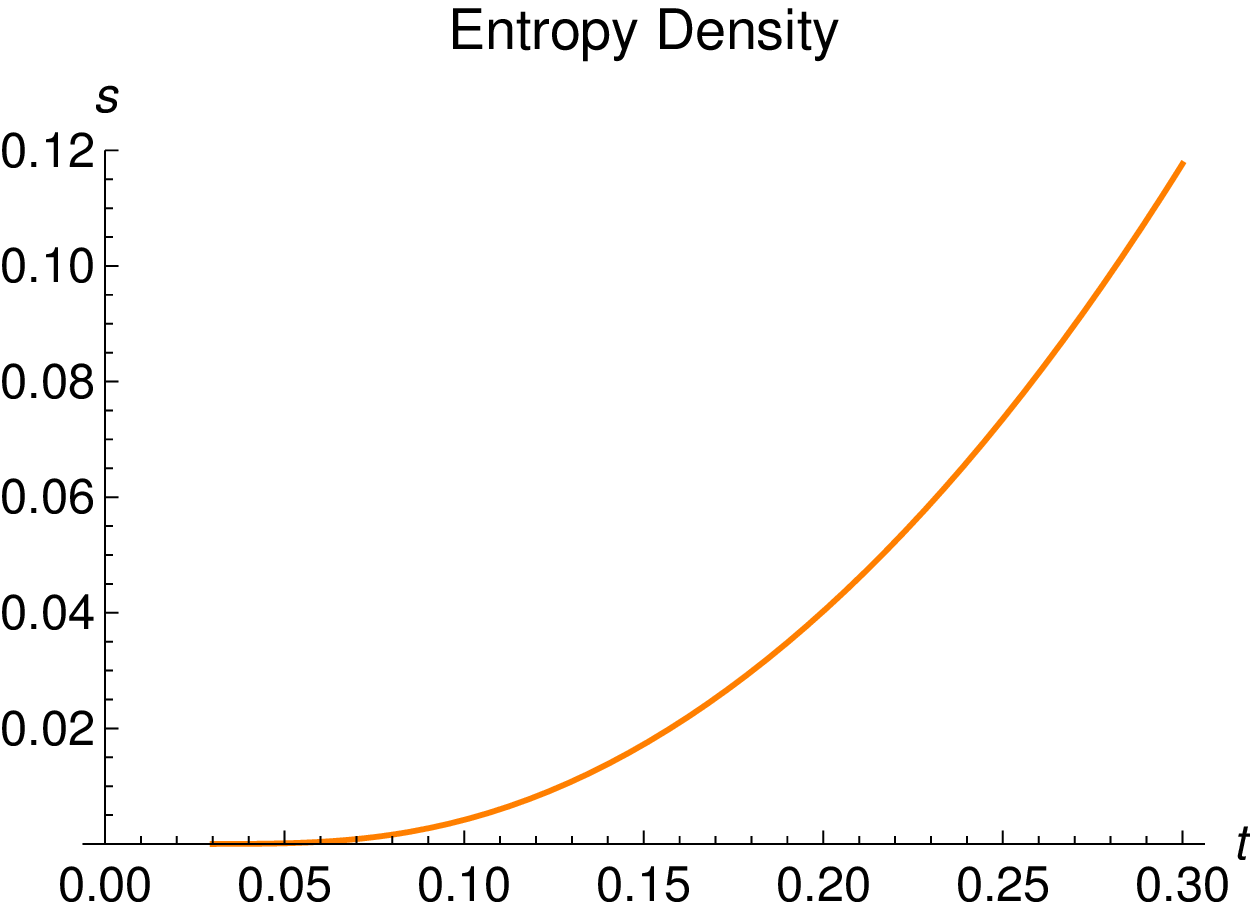}}
\end{center}
\caption{[Color online] Spin-$\frac{1}{2}$ square-lattice antiferromagnet in mutually orthogonal magnetic and staggered fields. LHS:
Temperature dependence of uniform (magenta) and staggered (brown) magnetization. RHS: Temperature dependence of entropy density. Upper
panel: $\{m,m_H\}=\{0.3,0.05\}$. Lower panel $\{m,m_H\}=\{0.3,0.2\}$.}
\label{figure9}
\end{figure}

In the absence of a magnetic field, the uniform magnetization is zero for arbitrary temperatures and arbitrary staggered field strength. In
presence of a magnetic field -- oriented perpendicular to the order parameter -- a uniform magnetization is induced that slightly increases
at more elevated temperatures according to Fig.~\ref{figure9}. But note that even in stronger magnetic fields corresponding to the point
$\{m,m_H\}=\{0.3,0.2\}$, the effect is very weak as compared to the uniform magnetization induced in the case of mutually parallel fields
(see LHS of Fig.~\ref{figure5}). Still, we are dealing with the analogous counterintuitive phenomenon because one would rather expect
thermal fluctuations to preclude creation of a uniform magnetization.

The staggered magnetization, on the other hand, decreases when temperature rises, as expected. The decrease is of the same magnitude as for
mutually parallel fields (see LHS of Fig.~\ref{figure5})\footnote{In order for the tiny effects in the uniform magnetization to become
visible, in Fig.~\ref{figure9} we have drawn up to $t=0.4$, whereas in Fig.~\ref{figure5} we plotted up to $t=0.3$.}. The sublattice
magnetizations $M_A$ and $|M_B|$ are the same for arbitrary temperature and for any point $\{m,m_H\}$ in parameter space, and just are half
of the staggered magnetization. As a consequence, due to this symmetric situation, no uniform magnetization can be created in the direction
of the order parameter: the uniform magnetization induced by the magnetic field points orthogonal to the staggered magnetization.

In the case of mutually parallel fields, the staggered magnetization served as an indicator for the total of aligned spins along the order
parameter axis. If the external fields are mutually orthogonal, in addition, we have to take into account the order created through the
uniform magnetization pointing into the direction of the magnetic field. Trying to correlate spin order with entropy, naively, the change
of entropy density with temperature is related to the decrease of the staggered magnetization and to the emergence of the uniform
magnetization as
\begin{equation}
\label{MagEntropyOrthogonal}
- \frac{\mbox{d} M_s}{\mbox{d} T} \propto \frac{\mbox{d} s}{\mbox{d} T} \, , \qquad
- \frac{\mbox{d} M}{\mbox{d} T} \propto \frac{\mbox{d} s}{\mbox{d} T} \, .
\end{equation}
Recall that $M_s$ and $M$ are the staggered and uniform magnetizations induced by finite temperature, i.e., reflecting the changes in these
quantities when temperature is raised from $t=0$ to $t \neq 0$.

Fig.~\ref{figure9} just evidences this scenario: the staggered magnetization decreases with temperature while the entropy density increases
-- the destruction of the arrangement of antialigned spins is the dominant effect. As the figure shows, the second contribution in
Eq.~(\ref{MagEntropyOrthogonal}) is negligible, such that the correlation is again between entropy density and staggered magnetization.
Note that the behavior of the entropy density -- as well as the staggered magnetization -- for both temperatures $t=\{0.2,0.4\}$ is
qualitatively the same: there is no "retardation effect" in the increase (decrease) of entropy density (staggered magnetization) in weaker
magnetic fields -- in contrast to the previous case of mutually parallel fields (see Fig.~\ref{figure5}), where the perturbation of spin
order in the stronger magnetic field was more pronounced.

A final remark concerns the validity domain of our effective analysis. As we have mentioned before, it applies to the domain where
temperature and external fields -- quantified by the parameters $t, m,m_H$ -- are small. In the case of mutually aligned magnetic and
staggered fields we furthermore have the stability criterion that we implemented by the condition $m > m_H + 0.1$. In general, as a
consequence of the Mermin-Wagner theorem, the staggered field in our effective analysis cannot be arbitrarily small at finite temperature
-- in particular it cannot be zero.\footnote{For a detailed discussion of why the effective approach fails in small staggered fields, the
interested reader may want to consult Figs. 2 and 3 of Ref.~\citep{Hof16b} and the information provided therein.} While the stability
criterion already guarantees that we never leave the domain where our effective analysis is valid, in the case of mutually orthogonal
fields we have been careful in the plots by not going below the value $m=0.1$ either. For this parameter region we are not aware of any
references -- neither theoretical nor experimental -- that have reported the counterintuitive effects we have described above.

However, in zero staggered field -- not captured by our study -- the initial increase of the magnetization with
temperature has been documented for the spin-$\frac{1}{2}$ square-lattice antiferromagnet in Refs.~\citep{FKLM92,San99}. Similar
counterintuitive order-disorder phenomena have also been seen in frustrated two-dimensional antiferromagnets (see Ref.~\citep{Sta15}) as
well as in antiferromagnetic spin chains and ladders. For example, according to Ref.~\citep{MHO07}, the magnetization of the critical
spin-$\frac{1}{2}$ antiferromagnetic chain initially increases as temperature rises. The effect has been confirmed experimentally (see,
e.g., Ref.~\citep{KSAHTLT15}). Then the behavior of the spin-$1$ antiferromagnetic chain is even more intriguing: the magnetization first
drops as temperature rises, goes through a minimum, and then grows. Finally, following Refs.~\citep{WY00,WOH01}, the magnetization of the
spin-$\frac{1}{2}$ two-leg spin ladder also presents a minimum.

\section{Conclusions}
\label{conclusions}

In the present systematic effective field theory investigation of two-dimensional bipartite antiferromagnets subjected to magnetic and
staggered fields, we focused on order-disorder phenomena by examining the behavior of the staggered, uniform and sublattice magnetizations
as well as the entropy density.

The first part dealt with the specific configuration of mutually aligned magnetic and staggered fields. In the entropy density at fixed
temperature we identified two opposite tendencies: (1) if the staggered field becomes stronger, the entropy decreases, and (2) if the
magnetic field becomes stronger, the entropy increases. Naively, the staggered field establishes spin order by enforcing antialignment of
the spins, whereas the magnetic field perturbs the antialigned array of spins. These observations are confirmed by the finite-temperature
behavior of the staggered, uniform and sublattice magnetizations. Remarkably, since the two sublattices of the bipartite antiferromagnet
are affected by the magnetic field in a nonsymmetric way, a finite-temperature uniform magnetization is induced that initially even grows
as temperature increases -- this outcome is rather counterintuitive.

In the second part we studied the configuration of mutually orthogonal magnetic and staggered fields. Overall, we find that the entropy
density at fixed temperature again diminishes when the staggered field grows, and that it increases when the magnetic field grows: likewise
the staggered field establishes spin order by enforcing antialignment of the spins, whereas the magnetic field perturbs this pattern. The
finite-temperature uniform magnetization that is induced perpendicular to the order parameter axis, slightly grows as temperature
increases, but here this counterintuitive phenomenon is tiny. The decrease of the staggered magnetization with temperature along the order
parameter axis is the dominant effect: microscopic order is destroyed.

While concrete plots referred to the spin-$\frac{1}{2}$ square-lattice antiferromagnet, our results and observations apply to any other
two-dimensional bipartite lattice. In this perspective the effective field theory analysis is universal -- details rely on the concrete
values of the spin stiffness and the zero-temperature staggered magnetization.

In conclusion, comparing the two different configurations of external fields antiferromagnets are subjected to, analogous characteristics
concern the emergence of counterintuitive phenomena, but the impact of the magnetic field in the case of mutually parallel fields is more
pronounced.

\begin{appendix}

\section{Kinematical Functions and Sunset Function for Antiferromagnetic Films in Mutually Orthogonal Magnetic and Staggered Fields}
\label{appendixA}

The dimensionless kinematical Bose functions
\begin{equation}
h^{I,{I\!I}}_0 = \frac{g^{I,{I\!I}}_0}{T^3} \, , \qquad h^{I,{I\!I}}_1 = \frac{g^{I,{I\!I}}_1}{T} \, , \qquad h^{I,{I\!I}}_2 = g^{I,{I\!I}}_1 \; T
\end{equation}
for magnon $I$ and magnon $I\!I$  are
\begin{eqnarray}
\label{BoseFunctions1}
h^{I}_0(H_s, H, T) & = & \frac{4 \pi^2 {(\sigma^2 + \sigma^2_H)}^{3/2}}{3}
- 2 \sqrt{\sigma^2 + \sigma^2_H} \; Li_2(e^{2 \pi \sqrt{\sigma^2 + \sigma^2_H} }) + \frac{1}{\pi} \; Li_3(e^{2 \pi \sqrt{\sigma^2 + \sigma^2_H} }) \nonumber \\
& & + 2 \pi (\sigma^2 +\sigma^2_H) \Big\{ \log(1- e^{-2 \pi \sqrt{\sigma^2 + \sigma^2_H}}) - \log(1- e^{2 \pi  \sqrt{\sigma^2 + \sigma^2_H} }) \Big\} \, ,
\nonumber \\
h^{I}_1(H_s, H, T) & = & - \frac{1}{2 \pi} \, \log \Big( 1 - e^{- 2 \pi\sqrt{\sigma^2 + \sigma^2_H}  } \Big) \, , \nonumber \\
h^{I}_2(H_s, H, T) & = & \frac{1}{8 \pi^2 \sqrt{\sigma^2 + \sigma^2_H}  \Big( e^{2 \pi \sqrt{\sigma^2 + \sigma^2_H} } - 1 \Big)} \, ,
\end{eqnarray}
and
\begin{eqnarray}
\label{BoseFunctions2}
h^{I\!I}_0(H_s, 0, T) & = & \frac{4 \pi^2 \sigma^3}{3} + 2 \pi \sigma^2  \Big\{ \log(1- e^{-2 \pi \sigma}) - \log(1- e^{2 \pi \sigma}) \Big\}
\nonumber \\
& & - 2 \sigma \; Li_2(e^{2 \pi \sigma}) + \frac{1}{\pi} \; Li_3(e^{2 \pi \sigma}) \, , \nonumber \\
h^{I\!I}_1(H_s, 0, T) & = & - \frac{1}{2 \pi} \, \log \Big( 1 - e^{- 2 \pi \sigma} \Big) \, , \nonumber \\
h^{I\!I}_2(H_s, 0, T) & = & \frac{1}{8 \pi^2 \sigma \Big( e^{2 \pi \sigma} - 1 \Big)} \, ,
\end{eqnarray}
respectively,  where $Li_2$ and $Li_3$ are polylogarithms. All quantities are expressed in terms of the two dimensionless parameters
$\sigma_H$ and $\sigma$,
\begin{equation}
\label{defSigmas}
\sigma_H = \frac{H}{2 \pi T} \, , \qquad \sigma = \frac{\sqrt{M_s H_s}}{2 \pi \sqrt{\rho_s} T} \, ,
\end{equation}
which in turn are related to $m_H$ and $m$ as
\begin{equation}
\sigma_H = \frac{\rho_s}{T} \, m_H \, , \qquad \sigma = \frac{\rho_s}{T} \, m \, .
\end{equation}
The dimensionless sunset function $s(\sigma,\sigma_H)$ is quite involved and we refer the reader to Ref.~\citep{Hof17}, where its
definition is given in (B14) and a plot is provided in Fig.~3.

\end{appendix}

\end{document}